\documentclass[twocolumn,showpacs,aps,pra,amsmath,amssymb,superscriptaddress]{revtex4-1}
\usepackage{bm,color,bbm}
\usepackage{hyperref, mathtools,graphicx,natbib}

\newcommand{\beq}{\begin{equation}}
\newcommand{\eeq}{\end{equation}}
\newcommand{\bqa}{\begin{eqnarray}}
\newcommand{\eqa}{\end{eqnarray}}
\newcommand{\nn}{\nonumber}

\newcommand{\smallfrac}[2]{\mbox{$\frac{#1}{#2}$}}

\newcommand{\half}{\smallfrac{1}{2}}

\definecolor{maroon}{rgb}{0.7,0,0}

\definecolor{ngreen}{rgb}{0.3,0.7,0.3}

\definecolor{golden}{rgb}{0.8,0.6,0.1}

\begin{document}
\title{Geometry of joint reality: device-independent steering and operational completeness}
\author{Michael J. W. Hall}
\affiliation{Department of Theoretical Physics, Research School of Physics, Australian National University, Canberra ACT 0200, Australia}
\affiliation{Centre for Quantum Dynamics, Griffith University, Brisbane, QLD 4111, Australia}
\author{\'Angel Rivas}
\affiliation{Departamento de F\'isica Te\'orica, Facultad de Ciencias F\'isicas, Universidad Complutense, 28040 Madrid, Spain}
\affiliation{CCS-Center for Computational Simulation, Campus de Montegancedo UPM, 28660 Boadilla del Monte, Spain}

\begin{abstract}
We look at what type of arguments can rule out the  joint reality  (or value definiteness)  of two observables of a physical system, such as a qubit, and give several strong yet simple no-go results based on assumptions typically weaker than those considered previously.  The first result uses simple geometry combined with a locality assumption to derive device-independent steering inequalities. These may also be regarded as ``conditional'' Bell inequalities,  are simpler in principle to test than standard Bell inequalities, and for two-qubit systems are related to properties of the quantum steering ellipsoid.  We also derive a Bell inequality from locality and a  one-sided reality assumption,  and demonstrate a close connection between device-independent steering and Bell nonlocality. Moreover, we obtain a no-go result without  the use of  locality or noncontextuality assumptions, based on similar geometry and an assumption that we call ``operational completeness''. The latter is related to, but strictly weaker than, preparation noncontextuality.   All arguments are given for finite statistics, without requiring any assumption that joint relative frequencies converge to some (unobservable) joint probability distribution. We also generalise a recent strong result of Pusey, for preparation noncontextuality, to the scenarios of device-independent steering and operational completeness.
\end{abstract}

\maketitle

\section{Introduction}

The question of whether and when real values can be attributed to quantum observables was raised by Einstein, Podolsky and Rosen (EPR) in 1935~\cite{epr}, and has continued to be debated,  in various forms, in the many decades thereafter. 
However, standard no-go results that rule out the joint reality of incompatible observables, based on Bell inequalities~\cite{bell, chsh,bellreview} and  Kochen-Specker arguments~\cite{ks,other,contreview}, do not apply to the simplest nontrivial quantum system, a single qubit.  One aim of this paper, therefore, is to give simple yet strong arguments that rule out joint reality even for this case. Further, our arguments are device-independent, being formulated for any two-valued observables of any physical system, whether or not it is described by quantum mechanics.

It is known that any such no-go argument must be based on an assumption of some sort. For example, two qubit observables $A,B=\pm1$  can be consistently assigned the joint real values $\alpha = {\rm sign} (\langle A\rangle_\psi-\lambda)$ and $\beta = {\rm sign} (\langle B\rangle_\psi-\lambda)$  prior to measurement,  for any pure state $\psi$, where $\lambda$ is a  random  variable uniformly distributed over $[-1,1]$~\cite{footex}. Thus, to meet the aims of the paper, there is a need to identify suitably simple and physically reasonable assumptions. Our no-go results are notable in that they are based on assumptions typically weaker than those found in the literature, and rely on very simple geometric arguments.

A further feature of the no-go results in this paper is the use of finite-counting arguments, so as to avoid any {\it a priori} assumption that  formal  joint probability distributions of incompatible observables  exist.   In particular, our results allow for scenarios in which the individual  relative  frequencies  of random variables converge (to corresponding  observable  probabilities) as the size of the ensemble increases, but where their joint relative frequency does not converge (see, e.g.,~\cite{rivaseprint} for an example). This is analogous to strong proofs of Bell inequalities in which only finite counting arguments are required~\cite{rivaseprint,mcdonald,gill}, thus allowing, for example, local hidden variable models based on properties of nonmeasurable sets~\cite{pitowsky} to be ruled out~\cite{mcdonald}. 

We begin in Sec.~\ref{sec:geom} by noting some very simple geometric constraints on the joint reality of two-valued observables, 
corresponding to  the elementary property that if two random variables have sufficiently large average values, then their product will have a large average value. These constraints form the basis of the main results of the paper, as indicated via simple qubit examples  (one of which supports a recent conjecture on the existence of formal quantum joint probability distributions~\cite{armen}).

Our first geometric no-go result is a device-independent steering inequality for the incompatibility of joint reality and locality in  Sec.~\ref{sec:steer}, which only assumes the reality of observables in a single spacetime region, and which can in principle  be tested with less detectors and/or assumptions than required in the standard  Bell nonlocality scenario.   In Sec.~\ref{sec:ellipsoid} we show that this inequality  is stronger than recent steering arguments given by Jevtic and Rudolph for qubit systems~\cite{jevtic}, in that it rules out the joint reality of arbitrary pairs rather than particular triples of projective qubit observables, and we relate the violation of the inequality to geometric properties of the quantum steering ellipsoid~\cite{sania, ellipsexp}.   In Sec.~\ref{sec:bell} we derive the Clauser-Horne-Shimony-Holt (CHSH) Bell inequalities from the assumptions of locality and {\it one-sided} reality,  and in Sec.~\ref{sec:bellsteer} we investigate the close connections between device-independent steering and Bell nonlocality, showing in particular that device-independent steering inequalities may be regarded as ``conditional'' Bell inequalities. We also reformulate a recent remarkable result of Pusey~\cite{pusey18}, to obtain a necessary and sufficient device-independent steering inequality for the CHSH scenario. Alternative necessary and sufficient inequalities are obtained in Appendix~\ref{appdiff}.

Section~\ref{sec:corr} is motivated by previous work that relates the joint reality of physical observables to the assumption of ``preparation noncontextuality''~\cite{spek05, contreview, natcommexp, pusey18, wolf},  i.e.,  to the requirement that operationally equivalent preparations have the same underlying distributions of any ontic variables. Preparation noncontextuality is, unlike Bell inequality and Kochen-Specker arguments, sufficient to rule out the joint reality of qubit observables~\cite{spek05,natcommexp}, and has a number of interesting implications~\cite{spek05, contreview, natcommexp, pusey18, wolf, spek08, spekPRL09,schmid}. 
In Sec.~\ref{sec:gen} we obtain a device-independent no-go result for joint reality based on a strictly weaker assumption, that we call ``operational completeness''---again using the simple geometric tools introduced in Sec.~\ref{sec:geom}.
In Sec.~\ref{sec:rob} we analyse the robustness of this result for the case of qubits. We directly compare preparation noncontextuality with operational completeness in Sec.~\ref{sec:compare}, and show that Pusey's recent result for preparation nontextuality~\cite{pusey18} can be strengthened to an analogous result for operational completeness.  

Conclusions are given  in Sec.~\ref{sec:con}.

\section{The geometry of joint reality}
\label{sec:geom}

We identify an observable of a physical system with a corresponding measurement procedure on that system. The `reality' of an observable captures the idea that its measurement reveals something already there, as follows. \\
{\bf Reality:} {\it An observable is defined to be a real property of a system if the measurement procedure corresponding to the observable acts to reveal a pre-existing value.}\\
Reality of an observable in this sense is also referred to  as `outcome determinism'~\cite{spek05} or `value definiteness'~\cite{jaeger}, and a quantum example was given in the Introduction.  While the reality or otherwise of a given observable might appear to be metaphysical in character, there are various physically testable implications for the {\it joint} reality of two or more observables~\cite{bellreview, natcommexp}. This paper obtains several rather simple but strong such implications.

In particular,  consider two observables $A$ and $B$ of some physical system (not necessarily quantum), having measurement outcomes labelled by $\alpha,\beta=\pm1$, respectively. For example, $A$ and $B$ might correspond to spin observables of a qubit, or to dichotomic observables of some classical or generalised probability theory. We will assume that $A$ and $B$ are jointly real  as per the above definition,   i.e., that they have well-defined real values $\alpha,\beta=\pm 1$, respectively, such that a measurement of either observable acts to reveal the corresponding value. 

\begin{figure}[!t] 
	\centering
	\includegraphics[width=0.55\textwidth]{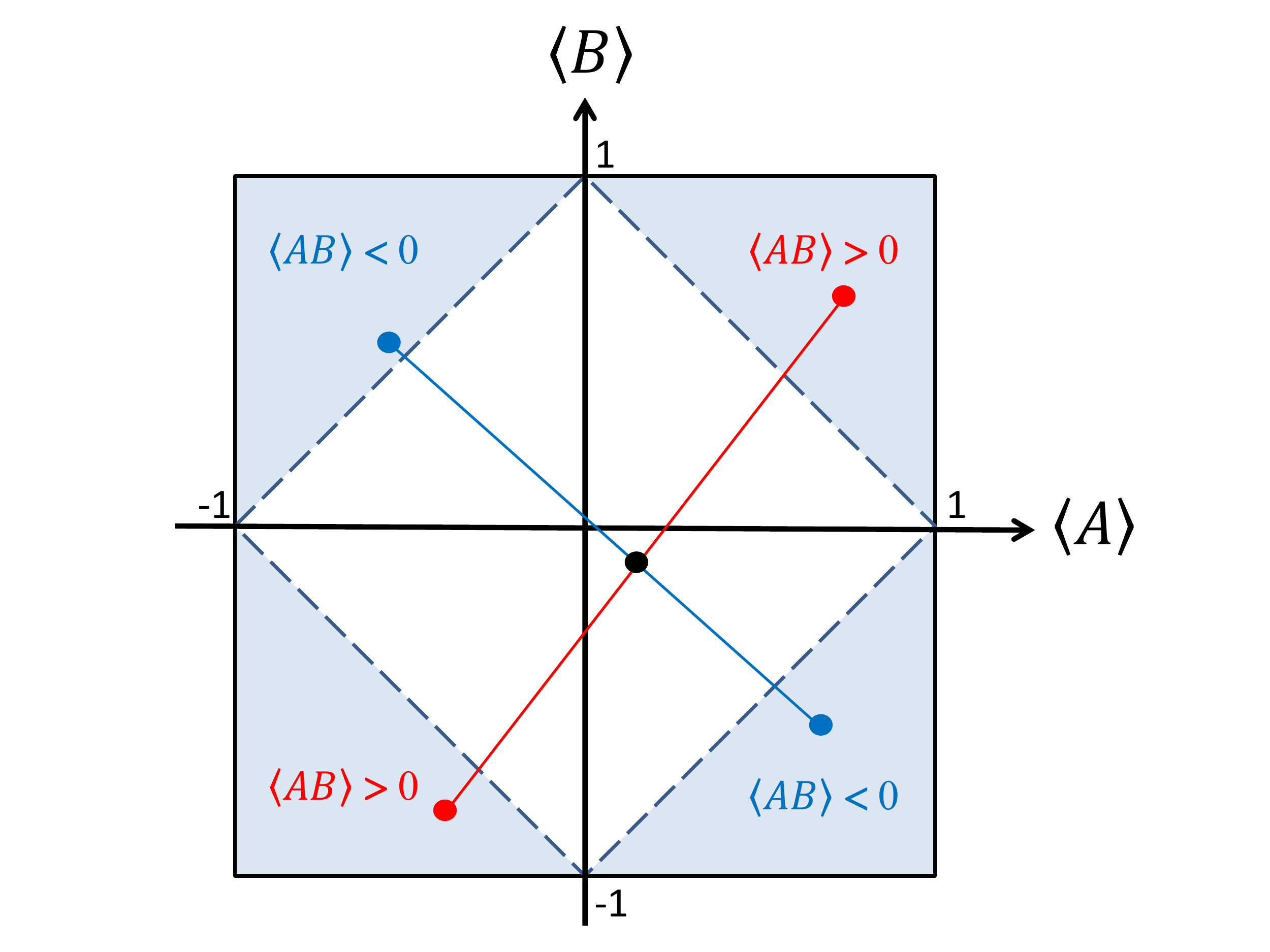}
	\caption{Simple constraints on joint reality. Let $A$ and $B$ be jointly real observables of an ensemble  of $N$ physical systems (not necessarily quantum), taking values $\pm1$. If  $\langle A\rangle$ and $\langle B\rangle$ are both sufficiently positive or negative then $\langle AB\rangle$ must be positive, as depicted by the shaded regions  in the upper right and lower left of the figure. Similarly, $\langle AB\rangle$ must be negative in the upper left and lower right shaded regions. The dashed boundary corresponds, via Eq.~(\ref{ident}), to the square $|\langle A\rangle|+|\langle B\rangle|=1$. It follows that any mixture of  ensembles corresponding to the two red dots must have $\langle AB\rangle>0$, while  any mixture of ensembles corresponding to the two blue dots must have $\langle AB\rangle<0$. Hence, ensembles having the values of $\langle A\rangle$ and $\langle B\rangle$ corresponding to the black dot can be equally well prepared with either strictly positive or strictly negative values of  $\langle AB\rangle$.  This immediately leads to robust no-go results under assumptions such as locality, as discussed in Sec.~\ref{sec:loc}, or  operational completeness or preparation noncontextuality, as discussed in Sec.~\ref{sec:corr}.
	}
	\label{semidi}
\end{figure}

For any ensemble of $N$ such systems, it follows that the ensemble averages $\langle A\rangle$, $\langle B\rangle$ and $\langle AB\rangle$ are well-defined via the relative frequencies $N(\alpha,\beta)/N$, where $N(\alpha,\beta)$ is the number of systems having real values $A=\alpha$ and $B=\beta$. 
We now make a simple observation: {\it if both $\langle A\rangle$ and $\langle B\rangle$ are sufficiently positive, then $\langle AB\rangle$ must also be positive.} More quantitatively, one has
\beq \label{ur}
\langle AB\rangle > 0 {\rm ~~for~~} \langle A\rangle+\langle B\rangle >1 ,
\eeq
corresponding to the upper right shaded region in Fig.~\ref{semidi}. This follows directly from the positivity of relative frequencies, using the identity~\cite{probid}
\beq \label{ident}
\frac{N(\alpha,\beta)}{N} =\frac{1}{4} \left[ 1 + \alpha\langle A\rangle + \beta\langle B\rangle + \alpha\beta\langle AB\rangle \right] \geq 0.
\eeq
In particular, choosing $\alpha=\beta=-1$ yields the inequality
\beq \label{ab+}
\langle AB\rangle \geq \langle A\rangle+\langle B\rangle -1 ,
\eeq
which immediately implies Eq.~(\ref{ur}). More generally, each choice of $\alpha$ and $\beta$ in Eq.~(\ref{ident}) yields a region of $\langle A \rangle$ and $\langle B\rangle$ values having a definite sign for $\langle AB\rangle$, corresponding to the four shaded regions in Fig.~\ref{semidi}.

It follows that any ensemble ${\cal E}_/$ lying on the red line in Fig.~\ref{semidi}, formed by a mixture of ensembles represented by the two red dots, must have $\langle AB\rangle_{{\cal E}_/}>0$, whereas any ensemble ${\cal E}_\backslash$ formed lying on the blue line in Fig.~\ref{semidi}, formed by a mixture of ensembles represented by the two blue dots, must have $\langle AB\rangle_{{\cal E}_\backslash}<0$. This simple fact immediately implies a general no-go result:  {\it the joint reality of $A$ and $B$ is incompatible with any assumption that equates the values of $\langle AB\rangle_{{\cal E}_/}$ and $\langle AB\rangle_{{\cal E}_\backslash}$ for two such mixtures} (e.g., for the case that ${\cal E}_/$ and ${\cal E}_\backslash$ have the same values of $\langle A\rangle$ and $\langle B\rangle$, corresponding to the black dot in Fig.~\ref{semidi}).

\begin{figure}[!t] 
	\centering
	\includegraphics[width=0.55\textwidth]{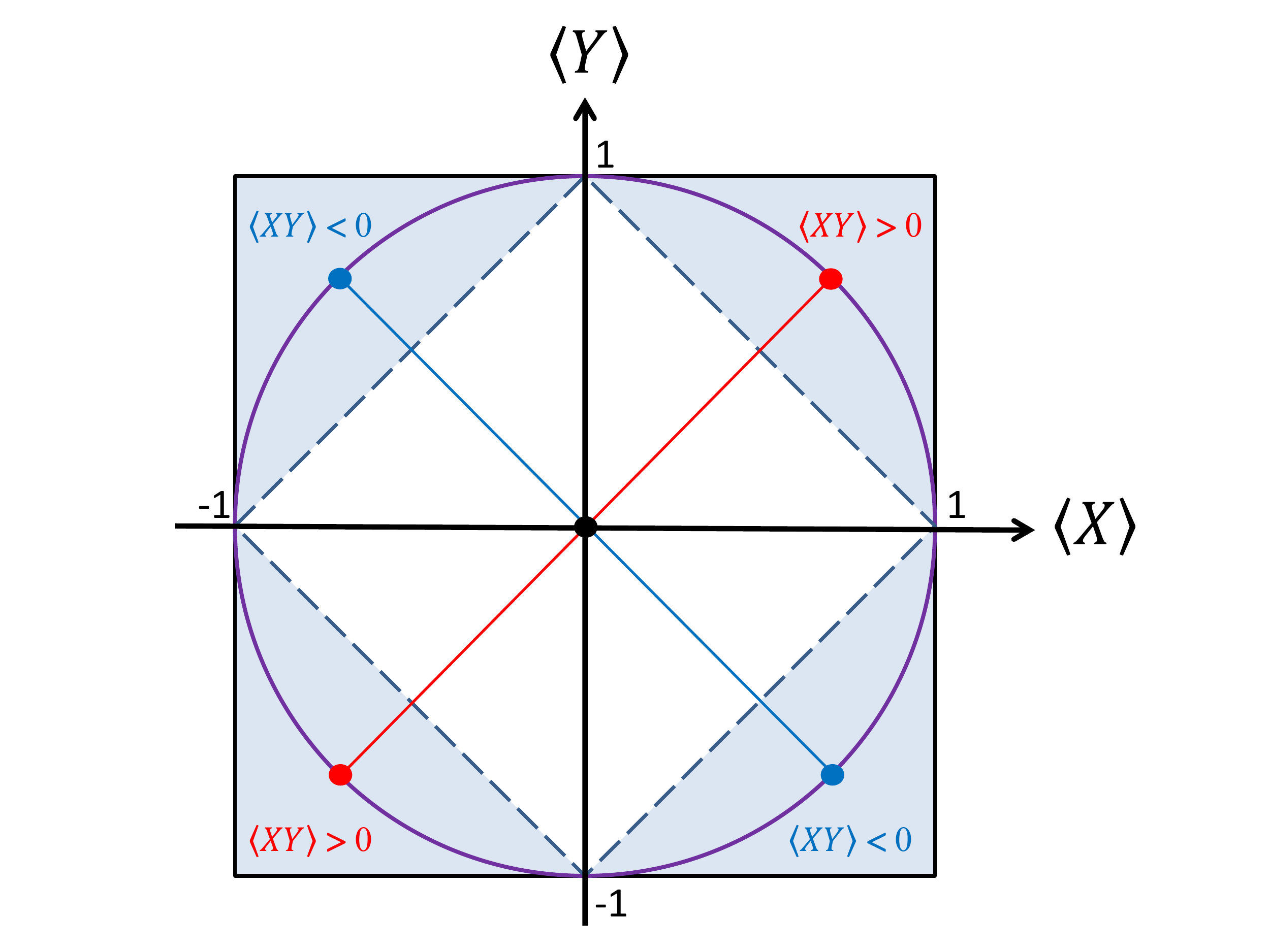}
	\caption{Joint reality vs locality for orthogonal qubit observables $X$ and $Y$. In quantum mechanics the range of possible $\langle X\rangle$ and $\langle Y\rangle$ values is restricted to the unit disc $\langle X\rangle^2+\langle Y\rangle^2\leq 1$ (purple solid curve). If the qubit is one member of a two-qubit singlet state, then one has $\langle X\rangle_{\cal E} = \langle Y\rangle_{\cal E} =0$ for an ensemble $\cal E$ of such states, corresponding to the black dot at the origin. A spin measurement on the second qubit, in the $\bm x+\bm y$ direction, will divide this ensemble into a mixture $\cal E_/$ of two subensembles represented by the red dots in the figure, implying as per Fig.~\ref{semidi} that $\langle XY\rangle_{{\cal E}_/} >0$. Similarly, a spin measurement in the $\bm x - \bm y$ direction will divide the ensemble into a mixture ${{\cal E}_\backslash}$ of the two subensembles represented by the blue dots, implying that $\langle XY\rangle_{{\cal E}_\backslash} <0$. 
		However, locality requires that the real values of $X$ and $Y$ are independent of which measurement is made on the second qubit, and hence that $\langle XY\rangle_{{\cal E}_/}=\langle XY\rangle_{{\cal E}_\backslash}$. Thus, locality is incompatible with the joint reality of $X$ and $Y$. Note that this result does not require any reality assumptions for the second qubit. It is generalised to device-independent steering inequalities (or conditional Bell inequalities) in Sec.~\ref{sec:loc}.
	}
	\label{xysimple}
\end{figure}

Suitable assumptions include locality, preparation noncontextuality, and a weakening of the latter which we call operational completeness. These lead to several simple device-independent no-go results, as discussed in Secs.~\ref{sec:loc} and~\ref{sec:corr}.
The flavour of these results is depicted in Fig.~\ref{xysimple} for the case of orthogonal qubit observables $X$ and $Y$,  corresponding to spin measurements in the $\bm x$ and $\bm y$ directions, respectively.

\begin{figure}[!t] 
	\centering
	\includegraphics[width=0.55\textwidth]{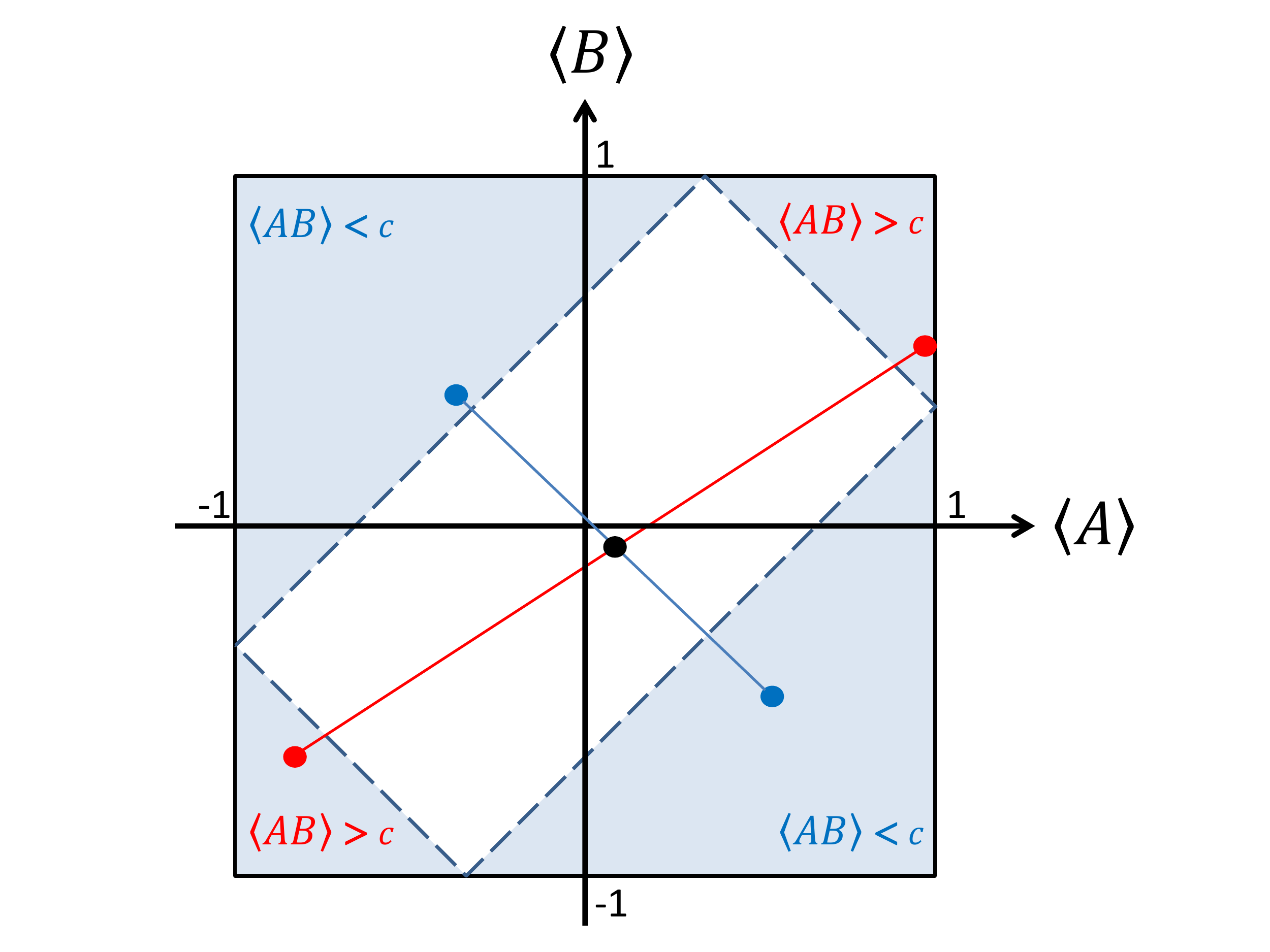}
	\caption{Simple generalised constraints on joint reality. Let $A$ and $B$ be jointly real observables of an ensemble  of $N$ physical systems (not necessarily quantum), taking values $\pm1$, and choose some $c\in(-1,1)$. 	The four shaded regions in the figure, corresponding to Eqs.~(\ref{rect1})--(\ref{rect4}), directly generalise those in Fig.~\ref{semidi} (which corresponds to the case $c=0$), with the dashed inner square replaced by a rectangle having vertices at $\pm(1, c),  \pm(c,1)$.  Analogously to Fig.~\ref{semidi}, any mixture of the ensembles corresponding to the two red dots must have $\langle AB\rangle> c$, whereas any mixture of the ensembles corresponding to the two blue dots must have $\langle AB\rangle <c$, implying that ensembles corresponding to the black dot can be equally well prepared with either strictly positive or strictly negative values of  $\langle AB\rangle-c$.  This immediately leads to the device-independent no-go results in Theorems~1 and~3.
	}
	\label{rectgen}
\end{figure}

We conclude this section by generalising the basic geometric result in Fig.~\ref{semidi}, to extend its useful range (and in particular to allow no-go results to be obtained for arbitrary noncommuting qubit observables).
This requires just a small modification of the observation made before Eq.~(\ref{ur}), as follows: {\it if both $\langle A\rangle$ and $\langle B\rangle$ are sufficiently large, then $\langle AB\rangle$ must also be large.} This is captured quantitatively, via Eq.~(\ref{ab+}), by the generalisation
\beq \label{rect1}
\langle AB\rangle > c \ {\rm ~~for~~} \langle A\rangle+\langle B\rangle >1+c 
\eeq
of Eq.~(\ref{ur}), for any $c\in(-1,1)$. It alternatively follows directly from the choice $\alpha=\beta=-1$ in identity~(\ref{ident}), while the remaining choices yield
\begin{align} \label{rect2}
&\langle AB\rangle > c \ {\rm ~~for~~} \langle A\rangle+\langle B\rangle<-1-c ,\\
&\langle AB\rangle < c \ {\rm ~~for~~} \langle A\rangle-\langle B\rangle >1-c  ,\\
&\langle AB\rangle < c \ {\rm ~~for~~} \langle A\rangle-\langle B\rangle <-1+c . \label{rect4}
\end{align}
These four equations correspond to the four shaded regions depicted in Fig.~\ref{rectgen}. In particular, the sign of $\langle AB\rangle - c$ is fixed to be positive in the upper right and lower left regions, and to be negative in the lower right and upper left regions. These regions, and the dashed rectangle that they exclude, reduce to those in Fig.~\ref{semidi} for $c=0$. 

While all four equations above are required for the device-independent results in Secs.~\ref{sec:loc} and~\ref{sec:corr}, we can already obtain a interesting result from Eqs.~(\ref{rect1}) and ~(\ref{rect2}) alone, under the assumption that quantum mechanics is valid. In particular, it has recently been conjectured there is no formal joint probability function $p(\alpha,\beta|\rho)$, for two quantum observables $A$ and $B$ and density operator $\rho$, under certain conditions~\cite{armen}. On any qubit subspace, these conditions require in particular that $\langle AB\rangle_{\rho_0}=\bm a\cdot\bm b$, for spin observables in nonparallel $\bm a$ and $\bm b$ directions and $\rho_0=\half \hat 1$ (e.g., orthogonal qubit observables $X$ and $Y$ are required to be uncorrelated for the maximally mixed state). Now, for such observables one can easily show that $|\langle A\rangle_{\rho_\pm} + \langle B\rangle_{\rho_\pm}|=|\bm a+\bm b|>1+\bm a\cdot\bm b$, where $\rho_{\pm}$ denotes a spin-up eigenstate in the $\pm(\bm a+ \bm b)$ direction. From Eqs.~(\ref{rect1}) and~(\ref{rect2}) it then follows that $\langle AB\rangle_{\rho_\pm}>\bm a\cdot\bm b$ (corresponding to the two red dots in Fig.~\ref{rectgen}, with $c=\bm a\cdot\bm b$). Thus, providing $\rho_0$ is prepared from an equally-weighted mixture of $\rho_\pm$, then $\langle AB\rangle_{\rho_0}>\bm a\cdot\bm b$, supporting the conjecture in Ref.~{\cite{armen}. Note no measurement or preparation noncontextuality assumption is required for this result.

\section{Locality and one-sided reality}
\label{sec:loc}

A very reasonable  physical  assumption, in the framework of relativity theory, is that  there are no  faster-than-light effects. We formalise this as follows.\\ 
{\bf Locality:} {\it  An operation carried out in some spacetime region cannot influence operations carried out in a spacelike separated region.}\\ 
 The combination of the reality and locality assumptions implies, in particular, that the real value of any observable that can be measured by an operation performed  within some spacetime region is unchanged by operations carried out in spacelike separated regions.  This combination, `local realism',  was the type of assumption that EPR had in mind when they considered making predictions``without in any way disturbing a system''~\cite{epr}, and was used by Bell in his derivation of the original Bell inequality~\cite{bell}.

It is well known the violation of a Bell inequality rules out the compatibility of locality and joint reality~\cite{bellreview} (modulo a measurement independence assumption, that measurement choices are uncorrelated with any variables that influence their outcomes~\cite{brans,hallfree}, which we will make throughout this paper).  However,  derivations of such inequalities typically use an  assumption equivalent to the  joint  reality of several observables in each of two spacelike separated regions~\cite{fineprl}. In this section, in contrast, we will obtain no-go results by only assuming {\it one-sided reality}, i.e., the joint reality of  particular  observables within a  {\it single} region of spacetime. In particular, no real pre-existing real values are assumed for measurements outside this region.

\subsection{Joint reality and device-independent steering}
\label{sec:steer}

 We consider again  an ensemble $\cal E$ of $N$ physical systems (not necessarily quantum), with two observables $A$ and $B$ that can be measured in some spacetime region and which have pre-existing real values $\alpha,\beta=\pm1$.  Under the above locality assumption, these values are undisturbed if an observer in a spacelike separated region makes one of two measurements, $M$ or $M'$ say, and records the result, $M=m$ or $M'=m'$ respectively. Note that we do {\it not} assume that $M$ or $M'$ have pre-existing real values before measurement.  

The measurement $M$ partitions $\cal E$ into a set of subensembles $\{ {\cal E}_m\}$,  where ${\cal E}_m$ comprises those systems   with outcome $M=m$. The alternative measurement $M'$ similarly partitions $\cal E$ into a set of subensembles $\{ {\cal E}'_{m'}\}$. Following Schr\"odinger, we say that the distant measurements {\it steer} each system into one of these subensembles~\cite{schr}. Since the real values of $A$ and $B$ are undisturbed by either measurement we can count the number of systems  in $\cal E$  having $A=\alpha$ and $B=\beta$ via either partition:
\beq \label{party}
N(\alpha,\beta|{\cal E}) = \sum_m N(\alpha,\beta|{\cal E}_m) =  \sum_{m'} N(\alpha,\beta|{\cal E}'_{m'}) .
\eeq
Denoting the number of systems in ${\cal E}_m$ and ${\cal E}'_{m'}$ by $N_m$ and $N'_{m'}$, respectively, it immediately follows that the average over $\cal E$ of any function $f(A,B)$ of the real values of $A$ and $B$, i.e., $\langle f(A,B)\rangle_{\cal E}:=N^{-1}\sum_{\alpha,\beta} f(\alpha,\beta) N(\alpha,\beta|{\cal E})$, can be decomposed into mixtures of the subensemble averages: 
\beq \label{local}
\langle f(A,B)\rangle_{\cal E}\! =\! \sum_m \frac{N_m}{N}\langle f(A,B)\rangle_{{\cal E}_m}\! =\! \sum_{m'} \frac{N'_{m'}}{N}\langle f(A,B)\rangle_{{\cal E}'_{m'}}.
\eeq
This is the basic observation that allows us to obtain the simple no-go results below.

For convenience, we will  now  restrict attention to the case where $M$ and $M'$ are two-valued measurements, with outcomes labelled by $m,m'=\pm1$. Thus, $M$ steers each system into one of  two  subensembles ${\cal E}_\pm$, and $M'$ into one of  two  subensembles ${\cal E}_\pm'$.  If the values of $\langle A\rangle$ and $\langle B\rangle$ are plotted as points on the $\langle A\rangle\langle B\rangle$-plane, for each of these subensembles, it follows from Eq.~(\ref{local}) that the points form a convex quadilateral, with diagonals intersecting at the point $(\langle A\rangle_{\cal E},\langle B\rangle_{\cal E})$.  Figures~\ref{semidi} and~\ref{rectgen} provide suggestive examples of such formations, with the red dots representing ensembles  ${\cal E}_\pm$, the blue dots representing ${\cal E}_\pm'$, and the black dot representing ${\cal E}$.

However, under the joint reality and locality assumptions, it turns out that the four subensembles cannot in fact occupy all four shaded regions depicted in Fig.~\ref{semidi} or~\ref{rectgen}, as this would simultaneously require $\langle AB\rangle_{\cal E}-c$ to be both strictly positive and strictly negative. In particular, we have a simple  device-independent no-go result.\\
{\bf Theorem~1:} {\it The joint reality of any two-valued observables $A,B=\pm1$ in some spacetime region is compatible with locality for a given ensemble $\cal E$ only if, for any $c\in (-1,1)$ and any  two-valued measurements $M$ and $M'$ made in a spacelike separated region, the inequality
	\begin{align} 
	\ell(c)  &:=\min\big\{~\langle A\rangle_{{\cal E}_+} +\langle B\rangle_{{\cal E}_+}-1-c, \nn\\ 
	&\qquad~~~-\langle A\rangle_{{\cal E}_-} -\langle B\rangle_{{\cal E}_-}-1-c,\nn\\
	&\qquad~~~~~~  \langle A\rangle_{{\cal E}'_+} -\langle B\rangle_{{\cal E}'_+} -1+c,\nn\\ 
	&\qquad~~~ -\langle A\rangle_{{\cal E}'_-} +\langle B\rangle_{{\cal E}'_-}  -1+c~ \big\}\nn\\  
	&\leq 0  \label{th1}
	\end{align}
	holds for the steered subensembles ${\cal E}_\pm$ and ${\cal E}_\pm'$ corresponding to $M$ and $M'$, respectively. 
}\\
{\it Proof:} We proceed by contradiction. Suppose that $A$ and $B$ are jointly real, the locality assumption is satisfied, and that there are distant measurements $M, M'$ such that $\ell(c)>0$ for some $c\in(-1,1)$. Thus, each of the four expressions on the right hand side of Eq.~(\ref{th1}) is strictly positive, implying via Eqs.~(\ref{rect1})--(\ref{rect4}) that $\langle AB\rangle_{{\cal E}_\pm}>c$ and $\langle AB\rangle_{{\cal E}'_\pm}<c$ ( and  that the steered subensembles occupy the four shaded regions of Fig.~\ref{rectgen}). Hence, any mixture of the pair ${\cal E}_\pm$ must have $\langle AB\rangle>c$, while any mixture of the pair ${\cal E}_\pm'$ must have $\langle AB\rangle<c$. But locality implies that $\langle AB\rangle_{\cal E}$ can be expressed as a mixture of either pair, as per Eq.~(\ref{local}), yielding the desired contradiction (since $\langle AB\rangle_{\cal E}$ cannot be both  greater than and less than $c$). $\blacksquare$


A simple application of Theorem~1 is provided by the case depicted in Fig.~\ref{xysimple} for orthogonal qubit observables $X$ and $Y$, where $\cal E$ is an ensemble of two-qubit singlet states, and measurements $M$ and $M'$ correspond to measurements on the second qubit in the $\bm x+\bm y$ and $\bm x - \bm y$ directions.  For this case the subensembles ${\cal E}_\pm$ and ${\cal E}_\pm'$ are depicted by the red and blue dots in Fig.~\ref{xysimple}, corresponding to the Bloch vectors $\bm n_\pm=\pm(1,1,0)/\sqrt{2}$ and $\bm n'_\pm=\pm(1,-1,0)/\sqrt{2}$, which yields the violation
\beq \label{mmax}
\ell(0)=\sqrt{2}-1>0 .
\eeq
of Eq.~(\ref{th1}) for $c=0$. Hence, the joint reality of $X$ and $Y$ is incompatible with locality. Further results for qubits are given in the next subsection.

Note that the subensembles in Fig~\ref{xysimple} lie at the maximum possible distance  from the dashed square, $1-1/\sqrt{2}$, that is allowed by quantum mechanics. More generally, a value of $\ell(c)>0$ in Eq.~(\ref{th1}) of Theorem~1 corresponds, geometrically, to the four subensembles ${\cal E}_\pm, {\cal E}_\pm'$ occupying the four shaded regions in Fig.~\ref{rectgen}, and having a minimum distance of $\ell(c)/\sqrt{2}$ from the dashed rectangular region.

The main attractions of Theorem~1, in addition to its geometric simplicity,  are that (i) it relies on the physically very reasonable assumption of locality; (ii) it is both device-independent and theory-independent, i.e., it applies to any ensemble of physical systems whether or not they are described by quantum mechanics or some other theory;  and (iii) it  can have  an experimental advantage in comparison to the CHSH Bell inequality,  depending on the methods and assumptions used,  due to the averages in Eq.~(\ref{th1}) being conditioned on subensembles  rather than the full ensemble.

As a simple example of the latter point, consider a measurement of 
$\ell(c)$ for an ensemble of unheralded entangled polarisation qubits, using a single ideal photon detector on each side. In this scenario, the firing of the detector on the steering side is used to herald a chosen value for $M$ or $M'$, i.e., a corresponding subensemble $\cal E_\pm$ or $\cal E'_\pm$, with the detector being placed in the appropriate polarisation path on each run. Upon heralding, the detector on the steered side is used to determine the corresponding value of either $A$ or $B$ (note that since this second detector is ideal, it can always be placed in the +1 polarisation path, with a non-firing of the detector identified with a -1 result). The conditional expectations appearing in Eq.~(\ref{th1}) for $\ell(c)$, i.e.,  $\langle A\rangle_{\cal E_\pm}$ etc., can then be estimated from a long series of runs.  In contrast, a measurement of the CHSH Bell parameter cannot be made in this scenario (nor a test of no-signaling), without some further assumption to enable estimation of joint expectation values such as $\langle AM\rangle_{\cal E}$~\cite{bellreview} (see also Sec.~\ref{sec:bell} below). This is because, for an unheralded source, such joint expectation values can only be obtained from the above conditional expectation values if the steering probabilities $p(M=\pm1), p(M'=\pm1)$ can also be determined. But this is not possible without some further assumption---since, for example, the non-firing of both detectors on a given run cannot distinguish between (i)  polarisation values corresponding to the paths having no detector, and (ii) no entangled pair being produced. Suitable assumptions in this case include a constant source rate, with the detector on the steering side placed in each polarisation path for equal amounts of time~\cite{aspect, mitchell}, or an assumption that the statistics satisfy locality (see Sec.~\ref{sec:bellsteer}).

More generally, the number of detectors required in Bell inequality experiments depend on whether assumptions such as fair-sampling are made~\cite{bellreview}, as well as the quality of the source, the way the detectors are used, and the way the events are defined~\cite{aspect,mitchell,gisin,zeil}. It would be of interest in future work to compare these with the corresponding assumptions required for device-independent steering experiments.

A value  $\ell(c)>0$ certifies, modulo locality, that the values of $A$ and $B$ are not both predetermined.  This has practical significance for one-sided secure key distribution and randomness generation, similarly to the witnessing of EPR steering for quantum systems~\cite{steeringkey}. In particular no eavesdropper or adversary outside the local spacetime region has access to both values of $A$ and $B$ prior to their actual measurement.  We can therefore regard  a value $\ell(c) > 0$ in Eq.~(\ref{th1}) as a witness of {\it device-independent steering}. 

More generally, we  introduce the term  ``device-independent steering'' 
as corresponding to the inconsistency of the joint reality of a set of system observables  $A,B,C\dots$ with locality, under steering  of an ensemble $\cal E$  by a set of remote measurements $M, M', M'',\dots$, without any assumptions on the working of  preparation and  measurement devices. Such an inconsistency is evidenced by the incompatibility of the measured statistics of $A,B,C\dots$, for each steered subensemble  ${\cal E}_m, {\cal E}_{m'}', {\cal E}_{m''}'',\dots$, with the property that their joint relative frequencies must be positive. For example, Eq.~(\ref{th1}) of Theorem~1 corresponds, via Eqs.~(\ref{rect1})--(\ref{rect4}), to the positivity of joint relative frequencies as per Eq.~(\ref{ident}). 

It follows that device-independent steering, as defined above, differs from EPR steering. For example, the latter further requires that the system observables have well-characterised quantum descriptions~\cite{eprsteering}. While device-independent steering is also conceptually distinct from Bell nonlocality~\cite{bellreview}, there is in fact a very close connection. In particular, as will be shown in Sec.~\ref{sec:bellsteer}, device-independent steering inequalities such as Eq.~(\ref{th1}) may be reinterpreted as ``conditional Bell inequalities.''

\subsection{Examples: qubits and steering ellipsoids}
\label{sec:ellipsoid}


	%
\begin{figure}[!t] 
	\centering
	\includegraphics[width=0.55\textwidth]{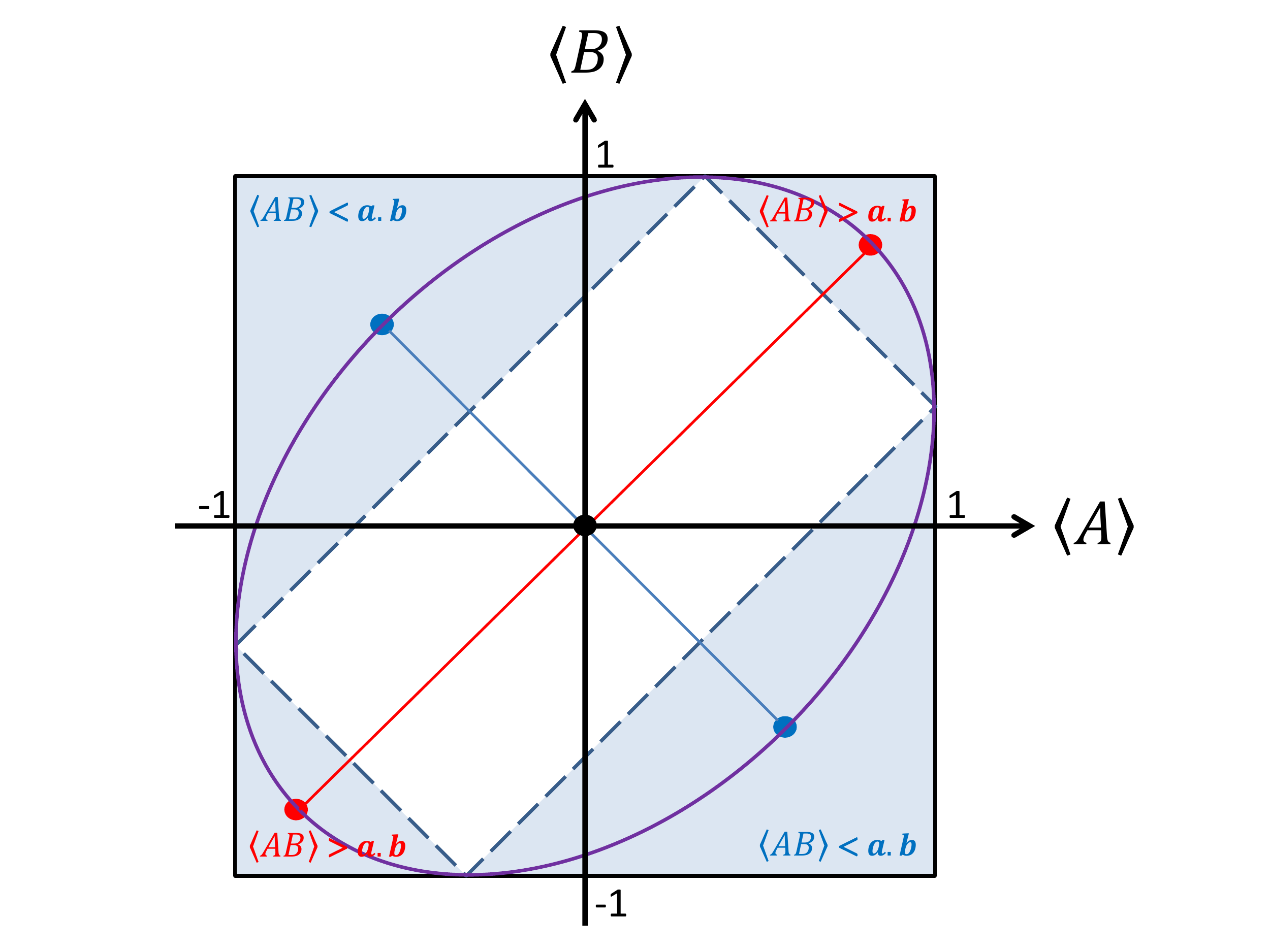}
	\caption{Joint reality vs locality for qubit observables  $A$ and $B$  represented by operators  $\hat A= \hat{\bm \sigma}\cdot\bm a$ and $\hat B= \hat{\bm \sigma}\cdot\bm b$. The four shaded regions in the figure correspond to the choice $c=\bm a\cdot \bm b$ in Fig.~\ref{rectgen}. For this choice, the range of possible $\langle A\rangle$ and $\langle B\rangle$ values is bounded by an ellipse as per Eq.~(\ref{ellipse}) (solid purple curve), which circumscribes the dashed inner rectangle. Hence, similarly to Fig.~\ref{xysimple} (which corresponds to the case of orthogonal qubit observables, with $c=\bm a\cdot\bm b = 0$), the incompatibility of joint reality with locality can be demonstrated via Theorem~1 if four suitable subensembles can be selected within the shaded regions of the ellipse. The optimal subensembles, maximising $\ell(c)$, correspond to the four points of the ellipse furthest from the dashed rectangle in each region (red and blue dots),  achievable for a singlet state  via measurements $M$ and $M'$ on the second qubit in the $\bm a\pm \bm b$ directions.
	}
	\label{rect}
\end{figure}

To apply Theorem~1  when $A$ and $B$ correspond to  projective qubit observables $\hat A= \hat{\bm \sigma}\cdot\bm a$ and $\hat B= \hat{\bm \sigma}\cdot\bm b$, where $\hat{\bm \sigma}\equiv (\hat X,\hat Y,\hat Z)$ denotes the vector of Pauli spin operators and $\bm a$ and $\bm b$ are unit vectors, note that the range of possible values for $\langle A\rangle$ and $\langle B\rangle$ in this case is bounded by the ellipse 
\beq \label{ellipse}
\frac{ \left(\frac{\langle A\rangle+\langle B\rangle}{\sqrt{2}}\right)^2}{1+\bm a\cdot\bm b} + \frac{ \left(\frac{\langle A\rangle-\langle B\rangle}{\sqrt{2}}\right)^2}{1-\bm a\cdot\bm b} \leq 1,
\eeq
depicted in Fig.~\ref{rect}. This may be shown by choosing two orthogonal observables $X'$, $Y'$ in the plane spanned by $\bm a$ and $\bm b$, and rewriting the tight bound $\langle X'\rangle^2+\langle Y'\rangle^2\leq 1$ for such observables in terms of $\langle A\rangle$ and $\langle B\rangle$. Note that the ellipse is oriented at 45$^\circ$, with semimajor and semiminor axis lengths $\sqrt{1\pm\bm a\cdot\bm b}$. The inscribed rectangle corresponds to $c=\bm a\cdot\bm b$ in Fig.~\ref{rectgen}. 
This immediately leads to a corollary of Theorem~1 for qubit observables.
\\
{\bf Corollary~1:} {\it The joint reality of any two noncommuting projective qubit observables,  represented by  $\hat A= \hat{\bm \sigma}\cdot\bm a$ and $\hat B= \hat{\bm \sigma}\cdot\bm b$, is incompatible with locality.}\\
{\it Proof:} Consider an ensemble described by the singlet state of the qubit and a spacelike separated qubit. Any projective measurement on the distant qubit will, therefore, steer the local qubit state to one of two opposite states on the Bloch ball. Choosing $M$ and $M'$ to steer the qubit to states described by the unit Bloch vectors in the directions $\pm(\bm a+\bm b)$ and $\pm(\bm a-\bm b)$, respectively, gives subensembles ${\cal E}_\pm$ and ${\cal E}'_\pm$ represented by  points of the ellipse  in each shaded region furthest  from the inscribed rectangle in Fig.~\ref{rect} (red and blue dots). Hence, since this rectangle corresponds to the value $c=\bm a\cdot\bm b$ in Theorem~1, we have $\ell(\bm a\cdot\bm b)>0$ in Eq.~(\ref{th1}). $\blacksquare$

The corollary improves on the two no-go results obtained by Jevtic and Rudolph based on steering properties of two-qubit states~\cite{jevtic}, in that (i)~the latter require the assumption of joint reality for {\it three} observables rather than two (e.g, for the observables having respective eigenstates $|a\rangle, |b\rangle, |x\rangle$ in Fig.~1 of~\cite{jevtic}), and (ii)~the corollary may be tested in a device-independent manner.

As noted in the caption of Fig.~\ref{rect}, the optimal subensembles which maximise $\ell(c)=\ell(\bm a\cdot\bm b)$   correspond  to the points of the ellipse in Fig.~\ref{rect} furthest from the dashed rectangle, depicted by the red and blue dots.  It is straightforward to calculate the corresponding maximum value of $\ell(\bm a\cdot\bm b)$ using Eq.~(\ref{ellipse}), as
\begin{align}
\ell_{\max}(\bm a\cdot\bm b) &= \min\{ \sqrt{2(1+ \bm a\cdot\bm b)} - (1+ \bm a\cdot\bm b),\nn \\ 
&\ \ \ \ \ \ \ \ \ \ \ \sqrt{2(1- \bm a\cdot\bm b)} - (1-\bm a\cdot\bm b) \} \nn\\  \label{lmaxab}
&= \sqrt{2(1+ |\bm a\cdot\bm b|)} - (1+ |\bm a\cdot\bm b|) .
\end{align}
This is a monotonic decreasing function of $|\bm a\cdot \bm b|$, ranging from a highest  value of $\sqrt{2}-1$  for $\bm a\cdot \bm b=0$, as expected from Eq.~(\ref{mmax}) for orthogonal observables, to a lowest value of 0 for $|\bm a\cdot \bm b|=1$ (corresponding to compatible observables $A=\pm B$). The device-independent nature of Theorem~1 implies that one can similarly calculate maximum values of $\ell(c)$ for positive-operator-valued measure (POVM) qubit observables (see also Sec.~\ref{sec:rob}).

The proof of Corollary~1 relies on the example of a two-qubit singlet state (see also Figs.~\ref{xysimple} and~\ref{rect}). However, the general result is very robust, and can be demonstrated with a wide variety of two-qubit states. To show this, let ${\cal S}({\cal E})$ denote the set of subensembles that a given physical ensemble $\cal E$ (not necessarily quantum) can be steered to, via arbitrary measurements made in spacelike separated regions. It is an immediate geometric consequence of Eq.~(\ref{th1}) that  a  necessary condition for obtaining $\ell(c)>0$ in Theorem~1 is that the image of  ${\cal S}({\cal E})$ in the $\langle A\rangle\langle B\rangle$-plane overlaps all sides of the dashed rectangle in Fig.~\ref{rectgen}.
For the particular case of  a general two-qubit state $\rho$, it is known that the set of steered local states of the first qubit, ${\cal S}(\rho)$, forms an ellipsoid in the Bloch ball $\langle X\rangle^2+\langle Y\rangle^2+\langle Z\rangle^2\leq 1$, called the quantum steering ellipsoid~\cite{sania, ellipsexp}. It is, therefore, of interest to reformulate the above necessary condition in Bloch coordinates, so as to make a connection with the geometry of quantum steering ellipsoids. 

For this purpose it may be assumed, without significant loss of generality, that $\hat A=\hat X$ and $\hat B=\hat X\cos\theta + \hat Y\sin\theta$ for orthogonal observables $X$ and $Y$ and some $\theta\in(0,\pi)$. Thus, the $\langle A\rangle\langle B\rangle$-plane is mapped to the $\langle X\rangle\langle Y\rangle$-plane via 
\beq \label{map}
\langle A\rangle=\langle X\rangle,\qquad\langle B\rangle=\langle X\rangle \cos\theta+\langle Y\rangle\sin\theta.
\eeq
For example, choosing $\theta=\tfrac{\pi}{2}$, it follows that {\it $\ell(c)>0$ is possible for observables $X$ and $Y$ only if  the projection of the steering ellipsoid onto the $\langle X\rangle\langle Y\rangle$-plane overlaps all four sides of the dashed rectangle in Fig.~\ref{rectgen}.} 

More generally, under the mapping in Eq.~(\ref{map}), the rectangle $R(c)$ in Fig.~\ref{rectgen} is mapped to a rectangle $\tilde R(c,\theta)$ in the $\langle X\rangle\langle Y\rangle$-plane, having  vertices at $\pm(1,(c-\cos\theta)/\sin\theta), \pm(c,(1-c\cos\theta)/\sin\theta)$, and sides oriented at angles $\theta/2$ and $\theta/2+\pi/2$.  This allows us to obtain a necessary and sufficient geometric condition as follows.\\
{\bf Corollary 2:} {\it The incompatibility of joint reality and locality for two qubit observables  represented by  $\hat X$ and $\hat X\cos\theta+\hat Y\sin \theta$ may be demonstrated via $\ell(c)>0$ in Eq.~(\ref{th1}), using an entangled two-qubit state $\rho$, if and only if there  are two line segments, each connecting two boundary points of the quantum steering ellipsoid $S(\rho)$ and passing through $(\langle X\rangle_\rho,\langle Y\rangle_\rho,\langle Z\rangle_\rho)$, such that (i) the projection of the first segment onto the $\langle X\rangle\langle Y\rangle$-plane intersects opposite sides of the rectangle $\tilde R(c,\theta)$, and (ii)~the projection of the second segment intersects the other two sides of $\tilde R(c,\theta)$.}\\
{\it Proof:}
For any entangled state $\rho$ the steering ellipsoid satisfies the completeness property defined in Ref.~\cite{sania}, implying that for any two line segments connecting two boundary points of the quantum steering ellipsoid $S(\rho)$ and passing through $(\langle X\rangle_\rho,\langle Y\rangle_\rho,\langle Z\rangle_\rho)$, there exist measurements $M$ and $M'$ on the second qubit which respectively steer to subensembles ${\cal E}_\pm$ and ${\cal E}'_\pm$  corresponding to the endpoints of the first and second line segments~\cite{sania}.  But properties~(i) and~(ii) of the corollary are then equivalent to these subensembles projecting onto four points on the $\langle X\rangle\langle Y\rangle$-plane which fall outside the four sides of the rectangle $\tilde R(c,\theta)$, which is in turn equivalent via Eq.~(\ref{map}) to falling outside the four sides of rectangle $R(c)$ in the $\langle A\rangle\langle B\rangle$-plane, i.e., to $\ell(c)>0$.
$\blacksquare$

Noting that the steering ellipsoid must project into the unit disc $\langle X\rangle^2+\langle Y\rangle^2\leq1$, it follows from conditions~(i) and~(ii) of Corollary~2 that this disc  itself  must extend beyond the sides of the rectangle $\tilde R(c,\theta)$ if $\ell(c)>0$, i.e., the side lengths  of this rectangle  must be less than 2 units. These side lengths are given by $s_1=(1+c)\sec\frac{\theta}{2}$ in the $\theta/2$ direction and $s_2=(1-c)\,{\rm cosec}\,\frac{\theta}{2}$ in the $\theta/2+\pi/2$ direction, yielding the necessary condition 
\beq \label{nec1}
1-2\sin\frac{\theta}{2} < c  < 2\cos\frac{\theta}{2}-1 
\eeq
for $\ell(c)>0$. Further, it is easy to check that the vertices of $\tilde R(c,\theta)$ always lie outside the unit circle (corresponding to $s_1^2+s_2^2\geq 4$). Hence, in addition to Eq.~(\ref{nec1}). the projection of each line segment in Corollary~2 must pass through a corresponding rectangle formed by the intersection of the unit circle with $\tilde R(c,\theta)$, implying that the projection of $(\langle X\rangle_\rho,\langle Y\rangle_\rho,\langle Z\rangle_\rho)$ must lie within the intersection $I(c,\theta)$ of these two rectangles, i.e.,
\beq \label{nec2}
(\langle X\rangle_\rho,\langle Y\rangle_\rho) \in I(c,\theta) .
\eeq
Explicitly, $I(c,\theta)$ is the rectangle centred at the origin and having side lengths $\sqrt{4-s_1^2}$ and $\sqrt{4-s_2^2}$ in the $\theta/2$ and $\theta/2+\pi/2$ directions, respectively.  Note for any {\it pure} entangled state $|\psi\rangle$ that the steering ellipsoid is the entire Bloch ball~\cite{sania}. It follows for any such state with   $\langle Z\rangle_\psi=0$ that conditions~(\ref{nec1}) and~(\ref{nec2}) are not only necessary but sufficient for $\ell(c)>0$.

\subsection{Bell inequalities from one-sided reality}
\label{sec:bell}

As mentioned in the preamble of Sec.~\ref{sec:loc}, it is well known that violation of a Bell inequality rules out the compatibility of locality with the joint reality of observables measured in two or more spacetime regions~\cite{bellreview}. Fine showed, for example, that satisfying the CHSH Bell inequalities is equivalent to assuming a local realistic model of the statistics, with each observable in each spacetime region having real values~\cite{fineprl}. Here, we point out that the CHSH Bell inequalities can be derived by assuming the reality of the observables in {\it one}  spacetime region  only. 

In particular, consider the CHSH scenario for an ensemble $\cal E$ of $N$ systems, in which either of two observables $A$ and $B$ can be measured in one spacetime region and either of two observables $M$ and $M'$ can be measured in a spacelike separated region. We assume that $A$ and $B$ have pre-existing measurement results $\alpha,\beta=\pm1$, and label the measurement outcomes of $M$ and $M'$ by $m,m'=\pm 1$. Assuming locality then leads to Eq.~(\ref{party}) as before for the  steered  subensembles ${\cal E}_\pm, {\cal E}'_\pm$.
We can therefore define a (purely formal) {\it joint} probability distribution for the outcomes of $A$, $B$, $M$ and $M'$ as follows:
\beq \label{fine}
\wp(\alpha,\beta,m,m') := \frac{N(\alpha,\beta|{\cal E}_m)\,N(\alpha,\beta|{\cal E}'_{m'})}{N(\alpha,\beta|{\cal E})\,N},
\eeq
analogous to the construction used in Proposition~(1) of Fine~\cite{fineprl}. Note that it is positive, sums to unity via Eq.~(\ref{party}) and the identity $\sum_{\alpha,\beta}N(\alpha,\beta|{\cal E})=N$, and returns the correct  marginal distributions for the triples $(A,B,M),  (A,B,M')$.   But the CHSH Bell inequalities hold for any such joint probability distribution~\cite{fineprl}. For example, since $\alpha m+\alpha m'+\beta m-\beta m'=\pm 2$ for all possible outcomes, then
\beq \nn
-2 \leq \sum_{\alpha,\beta,m,m'} \wp(\alpha,\beta,m,m')\,(\alpha m+\alpha m'+\beta m-\beta m')\leq 2,
\eeq
yielding the  well known CHSH Bell inequality~\cite{chsh}
\beq \label{chsh}
\left|\langle AM\rangle_{\cal E} + \langle BM\rangle_{\cal E} +\langle AM'\rangle_{\cal E} -\langle BM'\rangle_{\cal E}\right|  \leq 2.
\eeq

Note that while the reality of $M$ and $M'$ has not been assumed in the above, an experimental test of such inequalities requires the that the outcome of either,  once measured, is not influenced by measurements of $A$ or $B$ in a spacelike separated region.  This follows from  our locality assumption.

\subsection{Device-independent steering vs Bell nonlocality}
\label{sec:bellsteer}

We now show that there is a close connection between device-independent steering and Bell nonlocality, and in particular that device-independent steering inequalities, such as Eq.~(\ref{th1}), may be regarded as ``conditional'' Bell inequalities.

First, recall from Sec.~\ref{sec:steer}, that device-independent steering concerns
the joint reality of a set of system observables  $A,B,C\dots$, for an ensemble $\cal E$ of $N$ systems steered by a set of remote measurements $M, M', M'',\dots$, with no assumptions made about the working of   preparation and  measurement devices. In particular, the inconsistency of such joint reality with locality corresponds to the incompatibility of the measured statistics of the observables with the property that the number of systems having particular values $\alpha, \beta, \gamma,\dots$ for $A,B,C,\dots$ must be positive for any steered subensemble ${\cal E}_m, {\cal E}_{m'}', {\cal E}_{m''}'',\dots$:
\beq \label{digen}
N(\alpha,\beta,\gamma,\dots|{\cal E}_m),~~ N(\alpha,\beta,\gamma,\dots|{\cal E'}_{m'}), ~\dots~ \geq 0.
\eeq
In contrast, Bell nonlocality in this scenario is equivalent to the inconsistency of locality with the existence of a joint probability distribution 
\beq \label{belljoint}
\wp(\alpha,\beta,\gamma,\dots,m,m',m'',\dots) \geq 0
\eeq
for all observables $A,B,C,\dots, M, M',M'',\dots$~\cite{fineprl, bellreview}. Thus, testing device-independent steering concerns {\it conditional} correlations between two spacelike separated regions, such as in Eq.~(\ref{th1}), whereas testing Bell nonlocality concerns {\it joint} correlations, such as in Eq.~(\ref{chsh}). As already noted in Sec.~\ref{sec:steer}, the former tests can in principle be experimentally simpler.

Remarkably, the two concepts are equivalent for finite ensembles.  To show this, first note that if  Eq.~(\ref{digen}) holds for the case of $k$ steering observables $M, M', M'', \dots$, then one can define a corresponding joint probability distribution
\begin{align}
\wp(\alpha,\beta,\gamma,&\dots,m,m',m'',\dots) :=\nn\\ ~&\frac{N(\alpha,\beta,\gamma\dots|{\cal E}_m)\, N(\alpha,\beta,\gamma\dots|{\cal E}'_{m'}) \dots }{N(\alpha,\beta,\gamma,\dots|{\cal E})^{k-1}\, N} .
\end{align}
This generalises Eq.~(\ref{fine}), and by construction satisfies Eq.~(\ref{belljoint}). Further, locality as per Eq.~(\ref{party}) (with $\alpha, \beta$ extended to $\alpha, \beta,\gamma,\dots$) ensures that it gives the correct marginal relative frequencies, e.g., with \begin{align}
\wp(\alpha,\beta,\gamma,\dots,m)&=\frac{N(\alpha,\beta,\gamma\dots|{\cal E}_m)}{N}\nn\\ \label{tricky}
&=\frac{N(\alpha,\beta,\gamma\dots|{\cal E}_m)}{N_m} \frac{N_m}{N} .
\end{align}
Thus, the combination of Eq.~(\ref{digen}) and locality implies Bell locality.  Conversely, if one is given a joint probability distribution as per Eq.~(\ref{belljoint}), then sampling it $N$ times yields a model for an ensemble $\cal E$ having jointly real values of $A,B,C, \dots$ (and of $M,M',M'',\dots$), and locality implies that the values of $A,B,C,\dots$ can be steered without disturbance by measurements of $M, M', M'', \dots$  into subensembles as per Eq.~(\ref{digen}). 

The above result shows that device-independent steering inequalities such as Eq.~(\ref{th1}) may be reinterpreted as {\it conditional} Bell inequalities. In particular, a value $\ell(c)>0$ in Eq.~(\ref{th1}),  for some $c$, certifies Bell nonlocality (and hence also certifies EPR steering~\cite{eprsteering}). Conversely, the CHSH Bell inequality certifies device-independent steering, as will also be seen more directly in Theorem~2 below. Note that the special case of a conditional Bell inequality  for  {\it three} spacelike separated regions, conditioned on (a fixed outcome of) a {\it single}  measurement in one of the regions, has been previously considered in quantum field theory~\cite{landau} and in loophole free tests of standard Bell inequalities~\cite{hanson,wiseman}. In contrast, Eq.~(\ref{th1}) requires only two spacelike separated regions, and is conditioned on  outcomes of two measurements in one of the regions.

Nevertheless, there remain some important and interesting differences  at the level of the {\it inequalities} corresponding to  device-independent steering  and Bell nonlocality. First, as noted  in Sec.~\ref{sec:steer},  the conditional nature of  device-independent steering inequalities  means that they can in principle be tested with fewer detectors  and/or assumptions.  Second, a more formal distinction arises for the case of infinite ensembles. In particular, a conditional inequality such as Eq.~(\ref{th1}) does not depend on the measurement or knowledge of the relative frequencies of steering measurement outcomes such as $N_m/N$ (these only appear in the locality condition~(\ref{local})), and hence device-independent steering inequalities do not require the assumption of the convergence of such relative frequencies to probabilities as $N$ increases. Third, it is a matter of experimental and technical interest as to whether the device-independent steering inequalities for a given scenario, such as Eq.~(\ref{th1}), which depend only on conditional subensemble averages and not on steering probabilities  such as $N_m/N$,  are sufficient to fully characterise Bell nonlocality for that scenario. Note that such sufficiency does not follow from the equivalence of device independent steering and Bell nonlocality demonstrated above, because the proof of this equivalence {\it does} require the steering probabilities $N_m/N$ (in addition to the conditional probabilities appearing in device-independent steering inequalities), as per Eq.~(\ref{tricky}).

We note that the third point above can in fact be settled affirmatively for the CHSH scenario. This follows by reformulating a strong result of Pusey for preparation noncontextuality~\cite{pusey18}, as a necessary and sufficient device-independent steering inequality for this scenario.  \\
{\bf Theorem~2:} {\it The joint reality of two two-valued observables $A,B=\pm1$ in some spacetime region is compatible with locality for a finite ensemble $\cal E$ if and only if  the inequality
	\begin{align} \label{iffsteer}
	\left|
	\begin{array}{cccc}
	\langle A\rangle_{{\cal E}_+}& \langle B\rangle_{{\cal E}_+}& p	\langle A\rangle_{{\cal E}_+}   +q\langle B\rangle_{{\cal E}_+} -1 &1\\
	\langle A\rangle_{{\cal E}_-}& \langle B\rangle_{{\cal E}_-}& r	\langle A\rangle_{{\cal E}_-}   +s\langle B\rangle_{{\cal E}_-} +1&1\\
	\langle A\rangle_{{\cal E}_+'}& \langle B\rangle_{{\cal E}_+'}& -r	\langle A\rangle_{{\cal E}_+'}   -s\langle B\rangle_{{\cal E}_+'} +1&1\\
	\langle A\rangle_{{\cal E}_-'}& \langle B\rangle_{{\cal E}_-'}&	-p\langle A\rangle_{{\cal E}_-'}    -q\langle B\rangle_{{\cal E}_-'} -1 &1\\
	\end{array}
	\right| \leq 0
	\end{align}
	holds, for all $p,q,r,s=\pm1$ satisfying $pqrs=-1$, and for all pairs of  subensembles ${\cal E}_\pm$ and ${\cal E}_\pm'$ steered by  two-valued  measurements $M$ and $M'$, respectively, made in a spacelike separated region,  where ${\cal E}_+, {\cal E}_+', {\cal E}_-, {\cal E}_-'$ form a convex quadrilateral in the $\langle A\rangle\langle B\rangle$-plane (with respective vertices in clockwise order),
}\\
{\it Proof:} It is shown in~\cite{pusey18} that Eq.~(\ref{iffsteer}) is satisfied, under the stated conditions, if and only if the eight CHSH Bell inequalities are satisfied for  measurements of $A$ or $B$ on one side and of $M$ or $M'$ on the other (note that our labelling ${\cal E}_+, {\cal E}_-, {\cal E}_+', {\cal E}_-'$ corresponds to the labelling ${\cal P}_0, {\cal P}_3, {\cal P}_1, {\cal P}_2$ in~\cite{pusey18}, and we have cyclically permuted the bottom three rows of the determinant in Eq.~(11) of~\cite{pusey18}). But,  as noted at the beginning of Sec.~\ref{sec:bell},  the latter inequalities, and hence Eq.~(\ref{iffsteer}), are satisfied if and only if there exists a hidden variable model that specifies deterministic real values for $A$ and $B$ (and indeed also for $M$ and $M'$) and satisfies locality~\cite{fineprl}. 
$\blacksquare$

Since the eight CHSH inequalities fully characterise Bell  locality  in the CHSH scenario~\cite{bellreview}, it follows from the proof of the above theorem that the device-independent steering inequality~(\ref{iffsteer}) is similarly a full characterisation of Bell  locality  for fixed $M$ and $M'$ (note that there are eight  corresponding  possible choices of $p,q,r,s$). This appears rather remarkable at first sight, in that inequality~(\ref{iffsteer}) only depends upon conditional subensemble expectations, in contrast to the CHSH inequality in Eq.~(\ref{chsh}).  However, a different characterisation of Bell CHSH nonlocality via device independent steering inequalities is given in Appendix~\ref{appdiff},   including a simple demonstration  that the steering probabilities (and hence the full joint correlations) can be  directly  obtained from the conditional correlations in the CHSH scenario,  under the assumption that locality is satisfied.  

The device-independent steering inequality~(\ref{th1}) in Theorem~1  is weaker than inequality~(\ref{iffsteer}) in Theorem~2. The main value of the  former  is its much greater simplicity, characterised by a transparent geometric derivation, and its linearity, which  suggests that it may be more robust to statistical errors than Theorem~2.  Moreover,  Theorem~1 is sufficient, by virtue of Corollary~1 in Sec.~\ref{sec:ellipsoid},  to directly show that all projective noncommuting qubit observables are incompatible with locality, and to easily find the corresponding optimal qubit ensembles~(see also Fig.~\ref{rect}). In contrast, a full proof of Theorem~2 requires wading through an ``algebraic quagmire'' to obtain Eq.~(11) of~\cite{pusey18}, and the highly nonlinear set of inequalities as per Eq.~(\ref{iffsteer})  are difficult to optimise for the case of qubit observables. 

It would be of interest to further examine the connections between these theorems, as well as to test both theorems experimentally.
Note that any such experimental test  will in practice use some preparation procedure $P$ to prepare two ensembles, ${\cal E}$ and $\tilde{\cal E}$, on which $M$ and $M'$ are perfomed respectively. Hence these ensembles will be statistically similar rather than identical, leading to errors from finite statistics similarly to standard Bell inequality tests.

\section{Device-independent no-go results without locality or noncontextuality}
\label{sec:corr}

The results of the previous section required consideration of two spacelike separated regions and a locality assumption. However, while locality is physically well motivated, it is also of interest to instead consider what type of assumptions are needed to rule out joint reality based on  physical operations  in a single region only, including for the case of measurements on a single qubit. As mentioned in the introduction, a known assumption of this type is preparation noncontextuality~\cite{spek05}. Here we formulate and prove device-independent results for joint reality based on a weaker assumption that we call ``operational completeness'', using geometric arguments similar to those of the previous sections.

\subsection{Joint reality and operational completeness}
\label{sec:gen}

Recall from the discussion of Fig.~\ref{semidi} in Sec.~\ref{sec:geom} that a simple no-go result for the the joint reality of two observables $A,B=\pm1$ can be obtained via any assumption that equates the values of $\langle AB\rangle$ for two particular mixtures of ensembles. In Sec.~\ref{sec:loc} the assumption of locality was used. Here we consider a rather different alternative.\\
{\bf Operational completeness:} {\it If two ensembles are operationally similar, then the joint relative frequencies of any two observables having real pre-existing values are approximately the same for each ensemble.}\\
Here ``operationally similar'' means that the statistics of all  measurable observables are approximately the same for each ensemble, up to errors arising from finite statistics, expected to be of order $O(N^{-1/2})$.  For qubits, for example, it corresponds to the ensembles being  well described by the same Bloch vector. Likewise, ``approximately the same'' means up to errors that become arbitrarily small as $N$ increases.

Operational completeness is a rather strong assumption: the information contained in the statistics of all measurable observables, for a given ensemble, is sufficient to fix, at least approximately, the joint statistics of any pre-existing real values. Nevertheless, the assumption is strictly weaker than that of preparation noncontextuality, used in Refs.{~\cite{spek05, contreview, natcommexp, pusey18, wolf}}, as will be discussed in Sec.~\ref{sec:compare}, making it of some intrinsic interest. Note that it is typically a theory-dependent notion, because operational similarity requires specifying some fixed set of measurable observables (although such a specification may, alternatively, arise on purely phenomenological grounds,  in which case `phenomenological' could replace `operational' in the definition).   However, it is device-independent, i.e., it does not rely on any physical details of preparation and measurement devices. Hence, it leads to a device-independent no-go result.

It is convenient, for formulating this result, to first define a {\it operational plane} of ensembles, for a given pair of two-valued observables $A$ and $B$, as any set of ensembles such that: (i)  any two members ${\cal E}, {\cal E}'$ are operationally similar if $\langle A\rangle_{\cal E}\approx\langle A\rangle_{{\cal E}'}$ and $\langle B\rangle_{\cal E}\approx\langle B\rangle_{{\cal E}'}$ (up to errors arising from finite statistics); and (ii) it is closed under mixtures. Thus, an operational plane is intrinsically two-dimensional in character: its members can be distinguished by measurement, up to statistical errors, via two quantities: the values of $\langle A\rangle$ and $\langle B\rangle$.

For example, for the case of two incompatible projective qubit observables  $A$ and $B$ represented by  $\hat A= \hat{\bm \sigma}\cdot\bm a$ and $\hat B= \hat{\bm \sigma}\cdot\bm b$, the set of ensembles described by Bloch vectors of the form $\bm n=u\bm a +v\bm b$ forms an operational plane, parameterised by $\langle A\rangle=u+v\bm a\cdot \bm b$ and $\langle B\rangle=v+u\bm a\cdot \bm b$. The intersection of the Bloch ball with any plane not parallel to $\bm a\times\bm b$ also forms an operational plane for $A$ and $B$. Note that operational planes have an intrinsic uncertainty or `thickness' in practice,  due to finite statistics. The experimental  construction of ensembles lying in operational planes is discussed, e.g., in Appendix~C of~\cite{natcommexp} and Sec.~VIII of~\cite{pusey18}.

To see how operational completeness  can lead  to simple no-go results, note that we may reinterpret  the unit disc in  Fig.~\ref{xysimple} as the operational plane $\langle Z\rangle=0$ for qubit ensembles. Operational completeness then implies, for any ensemble corresponding to the black dot, that the value of $\langle XY\rangle$  is fixed by the values of $\langle X\rangle$ and $\langle Y\rangle$. Moreover, the joint reality of $X$ and $Y$ implies that there is one such ensemble, formed by  an equal  mixture of the two red dots, for which $\langle XY\rangle>0$, and a second such ensemble, formed by  an equal  mixture of the two blue dots, for which $\langle XY\rangle<0$. Hence, operational completeness is incompatible with the joint reality of qubit observables $X$ and $Y$.

More generally, we have a device-independent result.
\\
{\bf Theorem~3:} {\it The joint reality of two two-valued observables $A$ and $B$ is compatible with operational completeness only if, for any $c\in (-1,1)$, 
	\begin{align}
	\ell(c)&=\min\left\{~\langle A\rangle_1 +\langle B\rangle_1-1-c, \right.\nn\\ 
	&\qquad~~~-\langle A\rangle_2 -\langle B\rangle_2-1-c,\nn\\
	&\qquad~~~~~~  \langle A\rangle_3-\langle B\rangle_3-1+c,\nn\\
	&\qquad~~~\left. -\langle A\rangle_4 +\langle B\rangle_4-1+c~\right\}\nn\\  
	&\leq 0  \label{th3}
	\end{align}
	holds (up to statistical errors) for all ensembles ${\cal E}_1, {\cal E}_2, {\cal E}_3, {\cal E}_4$ in any operational plane of $A$ and $B$. 
}\\
{\it Proof:} Inequality~(\ref{th3}) is violated only if  the inequalities in (\ref{rect1})--(\ref{rect4}) are satisfied for ensembles ${\cal E}_1, {\cal E}_2, {\cal E}_3, {\cal E}_4$, respectively. But this corresponds to these ensembles having values of $\langle A\rangle$ and $\langle B\rangle$ in each one of the shaded regions of Fig.~\ref{rectgen} (as exemplified by the red dots for ${\cal E}_1, {\cal E}_2$ and by the blue dots for ${\cal E}_3, {\cal E}_4$). As already observed following Eq.~(\ref{ab+}), the joint reality of $A$ and $B$ then implies  two  mixed  ensembles ${\cal E}_/, {\cal E}_\backslash$ can be formed  in the operational plane  that have the same values of $\langle A\rangle$ and $\langle B\rangle$ (exemplified by the black dot in Fig.~\ref{semidi}),  up to statistical errors,  with $\langle AB\rangle_{{\cal E}_/}>0$ and $\langle AB\rangle_{{\cal E}_\backslash}<0$, respectively. But these two mixtures are operationally similar by the definition of an operational plane, and so must have the same values of $\langle AB\rangle$ under operational completeness  (again up to statistical errors).  Hence, joint reality is incompatible with operational completeness as claimed.$\blacksquare$

Note that the same symbol, $\ell(c)$, has been used in Eqs.~(\ref{th1}) and (\ref{th3}) for notational convenience, corresponding to formally equating the labels ${\cal E}_1, {\cal E}_2, {\cal E}_3, {\cal E}_4$ with ${\cal E}_+, {\cal E}_-, {\cal E}'_+, {\cal E}'_-$, respectively. Note also that violation of inequality~(\ref{th3}) requires an operational plane that extends outside all four sides of the dashed rectangle in Fig.~\ref{rectgen}. In particular, it is precisely in this case that four suitable ensembles ${\cal E}_1, {\cal E}_2, {\cal E}_3, {\cal E}_4$, such as those corresponding to the red and blue dots in Fig.~\ref{rectgen}, can be prepared.  For the qubit example discussed just prior to Theorem~3, these ensembles are  described by the Bloch vectors $\bm n_1=(1,1,0)/\sqrt{2}, \bm n_2=(-1,-1,0)/\sqrt{2}, \bm n_3=(1,-1,0)/\sqrt{2}, \bm n_4=(-1,1,0)/\sqrt{2}$, respectively, lying in the operational plane $\langle Z\rangle=0$ (see also Fig.~\ref{XYcase}), yielding a value $\ell(0)=\sqrt{2}-1$ similarly to the related `locality' example in Eq.~(\ref{mmax}).
 
More generally, we have a corollary of Theorem~3 for qubit observables, analogous to Corollary~2 in Sec.~\ref{sec:ellipsoid}.
\\
{\bf Corollary~3:} {\it The joint reality of any two noncommuting projective qubit observables,  represented by  $\hat A= \hat{\bm \sigma}\cdot\bm a$ and $\hat B= \hat{\bm \sigma}\cdot\bm b$, is incompatible with operational completeness.}\\
{\it Proof:} The  intersection of the  plane spanned by $\bm a$ and $\bm b$  with  the Bloch ball is an operational plane for $A$ and $B$, corresponding to the region in the $\langle A\rangle\langle B\rangle$-plane bounded by the ellipse defined in Eq.~(\ref{ellipse}). Since this ellipse circumscribes the dashed rectangle in Fig.~\ref{rect} for the choice $c=\bm a\cdot\bm b$, one can always choose four qubit ensembles ${\cal E}_1, {\cal E}_2, {\cal E}_3, {\cal E}_4$ lying in the four shaded regions, i.e., with  $\ell(\bm a\cdot\bm b)>0$ in Eq.~(\ref{th3}). In particular, choosing the ensembles described by unit Bloch vectors in directions $\pm(\bm a+\bm b), \pm(\bm a-\bm b)$ yields a value of $\ell(c)=\ell_{\max}(\bm a\cdot\bm b)$ in Eq.~(\ref{lmaxab}). $\blacksquare$

\begin{figure}[!t] 
	\centering
	\includegraphics[width=0.55\textwidth]{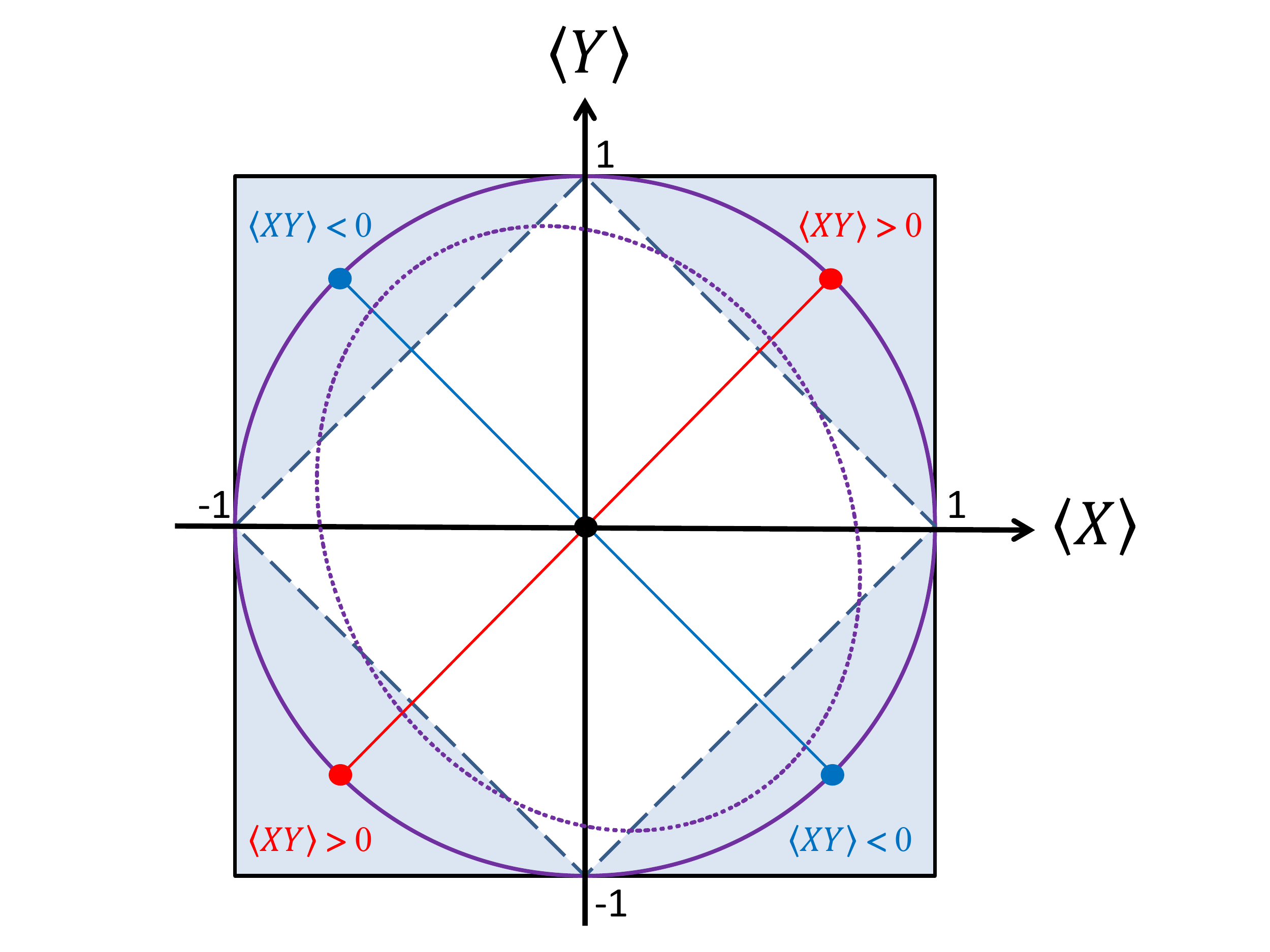}
	\caption{Joint reality vs operational completeness for orthogonal qubit observables $X$ and $Y$. Quantum mechanics restricts the range of possible $\langle X\rangle$ and $\langle Y\rangle$ values to the unit disc $\langle X\rangle^2+\langle Y\rangle^2\leq 1$ (purple solid curve). Thus, the four ensembles yielding the maximum possible violation of Eq.~(\ref{th3}) for $c=0$, i.e., having maximum possible distances from the dashed inner square, correspond to Bloch vectors in the $\pm\bm x\pm \bm y$ directions (red and blue dots).  These four ensembles lie in the operational plane $\langle Z\rangle =0$.  More generally, an arbitrary operational plane for $X$ and $Y$ has $\langle X\rangle$ and $\langle Y\rangle$ values restricted to the interior of an ellipse (exemplified by the purple dotted curve), as per the main text. In this case, inequality~(\ref{th3}) of Theorem~3 can be violated for $c=0$ if and only if the ellipse crosses all four sides of the dashed inner square. This is indeed the case for a large class of operational planes, as discussed in the main text, demonstrating the robustness of the theorem. Note that the ellipse plays an analogous role to the projection of the steering ellipsoid discussed in Sec.~\ref{sec:ellipsoid}.
	}
	\label{XYcase}
\end{figure}

\subsection{Robustness for qubits}
\label{sec:rob}

We will here exemplify the robustness of Theorem~3 for the special case of orthogonal qubit observables $X$ and $Y$ and $c=0$, and for related POVM observables. Similar results can be obtained for the general case.

First, as noted in Sec.~\ref{sec:gen}, an operational plane for observables $X$ and $Y$ corresponds to the intersection of the Bloch ball with any plane not parallel to the $\langle Z\rangle$ axis. Since the circle formed by such an intersection projects onto an ellipse in the equatorial plane, it follows that operational planes for $X$ and $Y$ are represented by the interiors of ellipses. Two examples are depicted in Fig.~\ref{XYcase} (purple curves), one of these being the unit disc (purple solid curve), corresponding to the special case of the operational plane $\langle Z\rangle=0$ discussed following Theorem~3. While this case leads to the maximum violation of Eq.~(\ref{th3}) for $c=0$, violations can in fact be achieved for a large class of operational planes, indicating the robustness of the theorem.

For example, the operational plane $\langle Z\rangle=d$ is parallel to the equatorial plane, and projects onto a disc of radius $\sqrt{1-d^2}$ centred at the origin in Fig.~\ref{XYcase}. Hence, it crosses all four sides of the inner square, corresponding to the existence of ensembles violating $\ell(0)\leq 0$ in Eq.~(\ref{th3}), for any $|d|< 1/\sqrt{2}$.  More generally, the boundary of an arbitrary operational plane for $X$ and $Y$, lying at distance $d$ from the centre of the Bloch ball and making an angle $\chi$ with the equatorial plane, projects onto an ellipse with semimajor and semiminor axis lengths $\sqrt{1-d^2}$ and $\sqrt{1-d^2}\cos\chi$, respectively (its orientation and centre are determined by the line through which the plane intersects the equatorial plane). Hence, a no-go result can be obtained for $c=0$ only if the width of this ellipse is larger than the side length of the dashed inner square, yielding the relatively mild necessary condition
\beq
\sqrt{1-d^2}\cos\chi > 1/\sqrt{2}  .
\eeq
This condition is also sufficient when the ellipse axes are parallel to the inner square.

Similar robustness considerations hold for observables sufficiently close to $X$ and $Y$, including POVM observables.  For example, the POVMs $X_\epsilon\equiv\{\hat X_\epsilon^\pm=\half(1\pm\epsilon \hat X)\}$, $Y_\epsilon\equiv\{\hat Y_\epsilon^\pm=\half(1\pm \epsilon \hat Y)\}$ for some $\epsilon\in[0,1]$, representing unbiased noisy measurements of $X$ and $Y$, satisfy $\langle X_\epsilon\rangle=\epsilon \langle X\rangle$ and $\langle Y_\epsilon\rangle=\epsilon \langle Y\rangle$.  Thus, their range of possible values corresponds to a circle of radius $\epsilon$ in Fig.~\ref{XYcase}. The incompatibility of joint reality of $X_\epsilon$ and $Y_\epsilon$ with operational completeness can therefore be demonstrated if this circle intersects all four sides of the dashed inner square, i.e., if  
\beq
\epsilon>1/\sqrt{2} .
\eeq   
Note that this is a tight result, as $X_\epsilon$ and $Y_\epsilon$ are known to be compatible, with a well-defined joint probability distribution, for $\epsilon\leq 1/\sqrt{2}$~\cite{busch,teiko}.   The content of Theorem~3 is thus also applicable to the joint reality of noisy qubit observables, as expected from its device-independent nature. 



\subsection{Operational completeness vs preparation noncontextuality}
\label{sec:compare}

 We now show that  the property of operational completeness in Sec.~\ref{sec:gen} is closely related to the property of preparation noncontextuality introduced by Spekkens~\cite{spek05}. A little preliminary groundwork is required to define the latter. Note first from  Bayes theorem  that the probability of measurement outcome $\alpha$, for a measurement $A$ on a system prepared via preparation procedure $P$, can  always  be written in the form
\beq 
p(\alpha|A,P) = \int d\lambda\, p(\alpha|A,\lambda,P)\, p(\lambda|A,P),
\eeq
where $\lambda$ represents any additional relevant information. An  {\it ontological model}, for a given set of measurement and preparation procedures, is then  said to exist when $P\rightarrow\lambda\rightarrow \alpha$ forms a Markov chain, i.e., when $p(\lambda|A,P)=p(\lambda|P)$ and $p(\alpha|A,\lambda,P)=p(\alpha|A,\lambda)$, implying that~\cite{spek05}
\beq \label{ontic}
p(\alpha|A,P) = \int d\lambda \,p(\alpha|A,\lambda) \,p(\lambda|P).
\eeq
In this case $\lambda$ is called an `ontic state', and is interpreted as carrying all relevant information about the system post-preparation and pre-measurement.  

Second, an ontological model is defined to be {\it preparation noncontextual} if any two operationally equivalent preparations $P$ and $P'$ in the given set are also equivalent at the ontic level~\cite{spek05,mnc}, i.e., if
\beq \label{pnc}
p(\alpha|A,P)\equiv p(\alpha|A,P') ~\forall A\implies
p(\lambda|P) \equiv p(\lambda|P') .
\eeq
Preparation noncontextuality  is a very strong assumption for ontological models. It rules out, by fiat, the possibility of being able to discriminate physical systems at some underlying level if they cannot be discriminated at an operational or phenomenological level. This is analogous to ruling out consideration of statistical mechanics on the basis of thermodynamic observations, or the existence of atoms based on the successes of fluid mechanics. Nevertheless, analysis of preparation noncontextuality  has led to a number of interesting implications~\cite{spek05, contreview, natcommexp, pusey18, wolf, spek08, spekPRL09,schmid}.

Of particular interest for us here is the following Proposition.\\
{\bf Proposition:} {\it Within the realm of ontological models, preparation noncontextuality implies operational completeness, but not vice versa}.\\
{\it Proof:} Note first that the pre-existing reality of two observables $A$ and $B$ is represented  within the realm of  ontological models by outcome determinism~\cite{spek05}, i.e, by
\beq \label{outdet}
p(\alpha|A,\lambda), ~p(\beta|B,\lambda) \in \{0,1\}.
\eeq
Defining the measurable set of ontic states $S^\alpha_A:=\{\lambda:p(\alpha|A,\lambda)=1\}$, and similarly for $S^\beta_B$, it follows from Eqs.~(\ref{ontic}) and~(\ref{outdet}), noting the intersection of two measurable sets is always a measurable set, that the joint probability of $A=\alpha$ and $B=\beta$ is well-defined for any preparation $P$: 
\beq \label{determ}
p(\alpha,\beta|P) = \int_{S^\alpha_A\cap S^\beta_B} d\lambda\, p(\lambda|P) .
\eeq
 Hence, if two preparations $P$ and $P'$ are operationally equivalent, then preparation noncontextuality as per Eq.~(\ref{pnc}) implies that
\beq \label{joint}
p(\alpha,\beta|P) =p(\alpha,\beta|P') .
\eeq
In particular, corresponding ensembles ${\cal E}$ and ${\cal E}'$ of $N$ systems prepared by these procedures will be operationally similar, and sample $p(\alpha,\beta|P)$ and $p(\alpha,\beta|P')$ respectively, implying that 
\beq \label{opcom}
N(\alpha,\beta|{\cal E})/N\approx N(\alpha,\beta|{\cal E}')/N
\eeq 
up to statistical errors. Thus operational completeness, as defined in Sec.~\ref{sec:gen}, holds for any ontological model that satisfies preparation noncontextuality, as claimed.

We show that the converse does not hold via a simple counterexample. Consider an ontological model that includes just two measurements $A$ and $B$ that have pre-existing real values $\alpha,\beta=\pm1$; just two preparations $P$ and $P'$ that are operationally equivalent; and ontic states $\lambda$ taking values on the unit circle, such that
\beq
S^\alpha_A=\{\lambda: \alpha\sin\lambda\geq 0\},~~ S^\beta_B=\{\lambda: \beta\cos\lambda\geq 0\},
\eeq
\beq
p(\lambda|P)=\pi^{-1}\sin^2\lambda,~~p(\lambda|P')=\pi^{-1}\cos^2\lambda .
\eeq 
Substitution into  Eq.~(\ref{ontic})  then yields $p(\alpha|A,P)=p(\alpha| A,  P')$ and $p(\beta|B,P)=p(\beta|B,P')$, consistent with the operational equivalence of $P$ and $P'$. Further, substibution into Eq.~(\ref{determ})  yields $p(\alpha,\beta|P)=\frac{1}{4}=p(\alpha,\beta|P')$. Hence, the real values of $A$ and $B$ satisfy operational completeness. Yet clearly $p(\lambda|P)\neq p(\lambda|P')$ and so the model is not preparation noncontextual. $\blacksquare$

The Proposition shows that operational completeness is a strictly weaker assumption than preparation noncontextuality. This reflects the fact the former only places constraints at the ensemble level, as per Eqs.~(\ref{joint}) and (\ref{opcom}) (see also Sec.~\ref{sec:gen}), whereas the latter directly constrains the underlying ontic level as per Eq.~(\ref{pnc}). 
 
Another  important  point of difference between operational completeness and preparation noncontextuality is that the former can be defined and exploited without making any assumption that joint relative frequencies of jointly real observables converge to some well-defined joint probability distribution. In contrast, preparation noncontextuality is only defined within the ambit of ontological models, and hence only within the ambit of a well-defined joint probability distribution for jointly real observables as per Eq.~(\ref{determ}). 

Finally, it is worth noting that results obtained using the assumption of preparation noncontextuality~\cite{spek05, contreview, natcommexp, pusey18, wolf, spek08, spekPRL09,schmid}  can, in many cases, be strengthened to obtain similar results using the weaker assumption of operational completeness (particularly for the case of outcome deterministic ontological models). In this sense preparation noncontextuality can often be replaced by operational completeness. As an example, we here generalise a remarkable result of Pusey~\cite{pusey18} for preparation noncontextuality.\\
{\bf Theorem~4:} {\it The joint reality of two two-valued observables $A$ and $B$, for any four ensembles ${\cal E}_1, {\cal E}_2, {\cal E}_3, {\cal E}_4$ forming a convex quadrilateral  in any operational plane of $A$ and $B$ (with respective vertices in clockwise order),  is compatible with operational completeness if and only if  the inequality
	\begin{align} \label{iff}
	\left|
	\begin{array}{cccc}
	\langle A\rangle_1& \langle B\rangle_1& p	\langle A\rangle_1   +q\langle B\rangle_1 -1 &1\\
	\langle A\rangle_2& \langle B\rangle_2& -r	\langle A\rangle_2   -s\langle B\rangle_2 +1&1\\
	\langle A\rangle_3& \langle B\rangle_3& r	\langle A\rangle_3   +s\langle B\rangle_3 +1&1\\
	\langle A\rangle_4& \langle B\rangle_4&	-p\langle A\rangle_4    -q\langle B\rangle_4 -1 &1\\
	\end{array}
	\right| \geq 0
	\end{align}
	holds (up to statistical errors) for all $p,q,r,s=\pm1$ satisfying $pqrs=-1$.
}

The proof of the theorem is rather long, in contrast to the analogous result in Theorem~2 of Sec.~\ref{sec:bellsteer} for the incompatility of joint reality and locality, and we defer it to Appendix~\ref{appth4}. It is guided by the proof of the corresponding result for preparation noncontextuality in~\cite{pusey18}, but with some expanded detail for clarity, and some technical differences due to using the weaker notion of operational completeness, and the concepts of joint reality and finite ensembles in place of ontological models.

Theorem~3 is weaker than Theorem~4 in the same way that Theorem~1 is weaker than Theorem~2. However, similarly to the discussion at the end of Sec.~\ref{sec:bellsteer}, the main value of Theorem~3  in comparison to Theorem~4  is the much greater simplicity and geometric nature of its derivation; its linear form; and its straightforward optimisation for qubit observables.  Note that Appendix~\ref{appdiff} provides an alternative necessary and sufficient condition to Theorem~4 for the incompatibility of joint reality and operational completeness.

\section{Conclusions}
\label{sec:con}

We have given a number of device-independent no-go results for the joint reality of two-valued observables (Theorems~1--4 and Corollaries~1--3), based on various assumptions that are weaker than those considered previously.  The results in Theorems~1 and~3 are of particular interest in that they are based directly on the very simple geometry of correlations depicted in Figs.~\ref{semidi} and~\ref{rectgen};  are transparent to optimise for qubits;  and show that the rather different concepts of locality and operational completeness  have  a common   geometric  underpinning. 

There are several avenues for future work that are suggested by the results. First, it would be worthwhile to experimentally test and compare the device-independent steering inequalities~(\ref{th1}) and~(\ref{iffsteer}), in Theorems~1 and~2 respectively; investigate the possibility of their direct application to one-sided cryptographic key distribution and randomness generation~\cite{eprsteering};  and study the effects of detector inefficiencies.  
  More generally, it would be of value to establish whether Bell nonlocality can be completely characterised via conditional Bell inequalities, similarly to the special case of the CHSH scenario discussed in Sec.~\ref{sec:bellsteer}  and~Appendix~\ref{appdiff}. 

Second, it would be reasonably straightforward  to carry out an experimental test of Theorems~3 and~4, concerning the compatibility of joint reality with operational completeness (such an experiment would also be able to test the inequality  for preparation noncontextuality  in~\cite{pusey18}). It would be technically similar to the experiment reported by Mazurek {\it et al.}~\cite{natcommexp} (see also Sec.~\ref{sec:gen}), but with the advantage of ruling out the joint reality of any two  noncommuting  qubit observables, modulo operational completeness, rather than only of particular sets of three such observables.

Third, analysis of the relationship between Theorems~1 and~2, or, analogously, between Theorems~3 and~4, may give some geometric insight into both the CHSH Bell inequalities and Pusey's result in~\cite{pusey18}  (see also  Appendix~\ref{appdiff}). 

Fourth, while all results have been obtained for finite statistics, thus avoiding an implicit assumption that unobservable joint relative frequencies must converge to joint probability distributions (see Introduction), it may be possible to formulate these results more sharply via strict limits on statistical errors, as Gill has done for the case of Bell inequalities~\cite{gill}.

Finally, it would be of interest to further investigate the relationship between operational completeness and preparation noncontextuality, including to what degree the former is able to substitute for the latter in various scenarios (in addition to Theorem~4 in Sec.~\ref{sec:compare}).   Note  that operational completeness does not fall within the general notion of contextuality formulated by Spekkens, i.e., that a property which holds at the operational level should hold at an underlying ontic level~\cite{spek05} (for example, there is no joint relative frequency of incompatible qubit observables  available  at an operational level). Nor does it correspond to noncontextuality of the type assumed by Kochen and Specker~\cite{ks,other}, since it only requires one way of measuring any given observable. However, perhaps it could be shown, for example, that the existence of an ontological model satisfying operational completeness implies the existence of a  second  ontological model,  making the same predictions, that satisfies   preparation noncontextuality.

\acknowledgments
 We thank Armen Allahverdyan for bringing the conjecture in Ref.~\cite{armen} to our attention, and Shuming Cheng and an anonymous referee for suggesting several valuable clarifications.  MH is grateful for the receipt of Foundational Questions Institute Mini-Grant FQXi-MGB-1728, which enabled collaboration in Madrid. AR is grateful to the Spanish MINECO grants FIS2015-67411, FIS2017-91460-EXP, the CAM research consortium QUITEMAD S2018/TCS-4342, and US Army Research Office through grant W911NF-14-1-0103 for partial financial support.

\appendix

\section{Necessary and sufficient conditions for joint reality applicable to the CHSH scenario}
\label{appdiff}

We return to the geometry of correlations in Fig.~\ref{semidi} and Sec.~\ref{sec:geom}, to obtain far stronger albeit less simple results than inequalities~(\ref{th1}) and~(\ref{th3}) in Theorems~1 and~3. These results are  equivalent to, but quite different in form from, inequalities~(\ref{iffsteer}) and~(\ref{iff}) in Theorems~2 and~4,  and may be  directly applied  to the CHSH scenario.  

First, choosing $\alpha=\beta$ in the fundamental positivity condition~(\ref{ident}) for joint reality yields the lower bound
\beq \label{lowerab}
\langle AB\rangle \geq  L:=  |\langle A\rangle +\langle B\rangle| -1 
\eeq
for $\langle AB\rangle$, generalising inequality~(\ref{ab+}). One similarly finds, choosing $\alpha=-\beta$,
the upper bound
\beq
\langle AB\rangle \leq  U:= 1- |\langle A\rangle -\langle B\rangle| .
\eeq
 Noting $L\leq U$ for any given values $|\langle A\rangle|, |\langle B\rangle|\leq 1$, it follows that one has $N(\alpha,\beta)\geq 0$ as per Eq.~(\ref{ident}) for any value $\langle AB\rangle\in [L,U]$. The above inequalities are therefore tight in this sense.

For a mixture $\cal E_/$ of two ensembles $\cal E_+$ and $\cal E_-$, with respective mixing weights $w_+$ and $w_-$,  it follows via Eq.~(\ref{lowerab}) and $w_++w_-=1$ that 
we have the tight lower bound
\begin{align} 
\langle AB\rangle_{\cal E_/} &= w_+ \langle AB\rangle_{\cal E_+} +
w_- \langle AB\rangle_{\cal E_-} \nn\\ 
&\geq  w_+ |\langle A\rangle_{\cal E_+} +\langle B\rangle_{\cal E_+}| +w_- |\langle A\rangle_{\cal E_-} +\langle B\rangle_{\cal E_-}| -1\nn\\
&  := L_/  \label{lower} 
\end{align} 
 (note this is stronger than the related lower bound $|\langle A\rangle_{\cal E_/}+\langle B\rangle_{\cal E_/}|-1$).  Similarly, for a mixture $\cal E_\backslash$ of two ensembles $\cal E'_+$ and $\cal E'_-$, with respective mixing weights $w'_+$ and $w'_-$, we have the tight upper bound
\begin{align}
\langle AB\rangle_{\cal E_\backslash} 
&\geq 1 -w'_+ |\langle A\rangle_{\cal E'_+} -\langle B\rangle_{\cal E'_+}| - w'_- |\langle A\rangle_{\cal E'_-} -\langle B\rangle_{\cal E'_-}| \nn\\
&  =:  U_\backslash .  \label{upper}
\end{align} 
One obtains similar lower and upper bounds  $L_\backslash$ and $U_/$  when the roles of $\cal E_/$ and $\cal E_\backslash$ are reversed.\\
{\bf Lemma~1:} {\it Any assumption equating the values of $\langle AB\rangle_{\cal E_/}$ and $\langle AB\rangle_{\cal E_\backslash}$ is compatible with  the  joint reality  of two observables $A, B=\pm1$  if and only if}
\begin{align}
&w_+ |\langle A\rangle_{\cal E_+} +\langle B\rangle_{\cal E_+}| +w_- |\langle A\rangle_{\cal E_-} +\langle B\rangle_{\cal E_-}| \nn \\
&~~+ w'_+ |\langle A\rangle_{\cal E'_+} -\langle B\rangle_{\cal E'_+}| + w'_- |\langle A\rangle_{\cal E'_-} -\langle B\rangle_{\cal E'_-}| \nn \\ \label{lemma1}
& \leq 2
\end{align}
{\it and}
\begin{align}
&w_+ |\langle A\rangle_{\cal E_+} -\langle B\rangle_{\cal E_+}| +w_- |\langle A\rangle_{\cal E_-} -\langle B\rangle_{\cal E_-}| \nn \\
&~~+ w'_+ |\langle A\rangle_{\cal E'_+} +\langle B\rangle_{\cal E'_+}| + w'_- |\langle A\rangle_{\cal E'_-} +\langle B\rangle_{\cal E'_-}| \nn \\ \label{lemma2}
& \leq 2 .
\end{align}\\
{\it Proof:} Equation~(\ref{lemma1}) corresponds to the requirement that the  tight  lower bound in Eq.~(\ref{lower}) is no greater than the  tight  upper bound in Eq.~(\ref{upper}),  i.e., to $L_/\leq U_\backslash$,  while Eq.~(\ref{lemma2}) corresponds to the case that the roles  of $\cal E_/$ and $\cal E_\backslash$  are reversed,  i.e., to $L_\backslash \leq U_/$. Thus, these equations are necessary for the equality of $\langle AB\rangle_{\cal E_/}$ and $\langle AB\rangle_{\cal E_\backslash}$. Conversely, noting that one trivially has $L_/\leq U_/$ and $L_\backslash\leq U_\backslash$,  Eqs.~(\ref{lemma1}) and~(\ref{lemma2}) are also sufficient for equality, since one can choose $\langle AB\rangle_{\cal E_/}= k = \langle AB\rangle_{\cal E_\backslash}$, compatible with joint reality, for any $k$ satisfying $\max\{L_/,L_\backslash\} \leq k\leq \min \{U_/,U_\backslash\}$.   $\blacksquare$

For example, suppose two-valued measurements $M$ and $M'$, made in a spacelike separated region from measurements of $A$ and $B$, steer some ensemble $\cal E$ to subensembles $\cal E_\pm$, $\cal E'_\pm$, respectively. If $\cal E_\pm$ correspond to the two red dots in Fig.~\ref{semidi}, and $\cal E'_\pm$ correspond to the two blue dots, then the locality assumption implies that 
\beq
\langle AB\rangle_{\cal E_/}=\langle AB\rangle_{\cal E_\backslash} =\langle AB\rangle_{\cal E}
\eeq
via Eq.~(\ref{local}),  with $\cal E_/$ and $\cal E_\backslash$ corresponding to the black dot.  Hence, Lemma~1 applies to this scenario.

Further, since Eqs.~(\ref{lemma1}) and~(\ref{lemma2}) are, by construction,  equivalent to the positivity of the joint relative frequencies $N(\alpha,\beta|\cal E_\pm)$ as per Eq.~(\ref{ident}), they are then  also  equivalent to Bell locality in the CHSH scenario as per the argument in Sec.~\ref{sec:bellsteer}. Thus, {\it joint reality is compatible with locality in the CHSH scenario if and only if Eqs.~(\ref{lemma1}) and~(\ref{lemma2}) hold}.  

  Since  the two Eqs.~(\ref{lemma1}) and~(\ref{lemma2}) are equivalent to the eight CHSH inequalities in the CHSH scenario,  it  is  of interest to write them down  explicitly in terms of joint correlations. To do so, note first that the first term in Eq.~(\ref{lemma1}) can be rewritten in this scenario as
\begin{align}
	w_+ \langle A\rangle_{\cal E_+}&= p(M=1)\langle A\rangle_{M=1}  =  p(M=1)\langle A\,\tfrac{1+M}{2}\rangle_{M=1}\nn\\
	&= p(M=1)\langle A\,\tfrac{1+M}{2}\rangle_{M=1}\nn\\
	&~~~~\qquad + p(M=-1)\langle A\,\tfrac{1+M}{2}\rangle_{M=-1}\nn\\
	&= \langle A\,\tfrac{1+M}{2}\rangle_{\cal E} = \half \langle AM\rangle_{\cal E} + \half\langle A\rangle_{\cal E}  .
\end{align}
Treating the other terms similarly then gives the equivalent form
\begin{align}
&|\langle AM\rangle_{\cal E}+\langle BM\rangle_{\cal E}+\langle A\rangle_{\cal E}+\langle B\rangle_{\cal E}|\nn\\
&~+ |\langle AM\rangle_{\cal E}+\langle BM\rangle_{\cal E}-\langle A\rangle_{\cal E}-\langle B\rangle_{\cal E}|\nn\\
&~+|\langle AM'\rangle_{\cal E}-\langle BM'\rangle_{\cal E}+\langle A\rangle_{\cal E}-\langle B\rangle_{\cal E}|\nn\\
&~+ |\langle AM'\rangle_{\cal E}+\langle BM'\rangle_{\cal E}-\langle A\rangle_{\cal E}+\langle B\rangle_{\cal E}|\nn\\
\label{altchsh}
&\leq 4
\end{align}
of Eq.~(\ref{lemma1}). The corresponding form of Eq.~(\ref{lemma2}) is obtained by swapping $M$ and $M'$.  The two same inequalities are also obtained under swapping the outcomes of any of $A,B,M,M'$, and under swapping of $A$ and $B$.

Clearly, Eqs.~(\ref{lemma1}) and~(\ref{lemma2}) do {\it not} have the form of device-independent steering inequalities (or conditional Bell inequalities) in the CHSH scenario, since they depend explicitly on the steering weights $w_\pm, w'_\pm$ and so correspond to {\it joint} correlation (or Bell) inequalities as per Eq.~(\ref{altchsh}). Nevertheless, they may be put in such a form, via a special property of the CHSH scenario, as follows.\\
{\bf Lemma~2:}
{\it The mixing weights $w_\pm, w'_\pm$ in Eqs.~(\ref{lemma1}) and~(\ref{lemma2}) of Lemma~1 may be evaluated in terms of conditional expectation values with respect to $\cal E_\pm, \cal E'_\pm$, if the assumption referred to in Lemma~1 further equates the values of $\langle A\rangle_{\cal E_/}$ and $\langle A\rangle_{\cal E_\backslash}$, and of $\langle B\rangle_{\cal E_/}$ and $\langle B\rangle_{\cal E_\backslash}$.}\\
{\it Proof:} Writing $w_\pm=\half(1\pm\delta), w'_\pm=\half(1\pm \delta')$ for convenience, it follows that the equations $\langle A\rangle_{\cal E_/}=\langle A\rangle_{\cal E_\backslash}$ and $\langle B\rangle_{\cal E_/}=\langle B\rangle_{\cal E_\backslash}$ may be rewritten as two linear equations for $\delta$ and $\delta'$,  which  may be  given  in matrix form as
\beq \label{l1}
\left(
\begin{array}{cc}
\langle A\rangle_{\cal E_+}-\langle A\rangle_{\cal E_-} &
-\langle A\rangle_{\cal E'_+}+\langle A\rangle_{\cal E'_-}\\
\langle B\rangle_{\cal E_+}-\langle B\rangle_{\cal E_-} &
-\langle B\rangle_{\cal E'_+}+\langle B\rangle_{\cal E'_-}
\end{array}
\right)
\left(
\begin{array}{c}
\delta	\\\delta'
\end{array}
\right)
=
\left(
\begin{array}{c}
f\\f'	
\end{array}
\right),
\eeq
where
\begin{align} \label{l2}
f&= -\langle A\rangle_{\cal E_+}-\langle A\rangle_{\cal E_-} + \langle A\rangle_{\cal E'_+}+\langle A\rangle_{\cal E'_-}\\
\label{l3}
f'&= -\langle B\rangle_{\cal E_+}-\langle B\rangle_{\cal E_-} + \langle B\rangle_{\cal E'_+}+\langle B\rangle_{\cal E'_-} .
\end{align}
 Solving  for $\delta$ and $\delta'$ yields expressions for the weights $w_\pm, w'_\pm$ in terms of the averages of $A$ and $B$ with respect to $\cal E_\pm, \cal E'_\pm$, as claimed. Note that the solution corresponds to the black dot in Fig.~\ref{semidi}. $\blacksquare$

For example, since the locality assumption satisfies the conditions of Lemma~2 via Eq.~(\ref{local}), Eqs.~(\ref{lemma1}) and~(\ref{lemma2}) can be rewritten as device-independent steering inequalities  under this assumption.  Further, since these two inequalities are necessary and sufficient, it follows that they are equivalent to the eight device-independent steering inequalities in Eq.~(\ref{iffsteer}) of Theorem~2.

Lemmas~1 and~2 also allow the standard CHSH inequality in Eq.~(\ref{chsh}) to be directly reformulated as a nonlinear device-independent steering inequality, similarly to Eq.~(\ref{iffsteer}) of Theorem~2.  First, note that the first term of the inequality can be rewritten as
\begin{align}
\langle A M\rangle_{\cal E} &= p(M=1)\langle A\rangle_{M=1} - p(M=-1)\langle A\rangle_{M=-1} \nn\\
&=w_+\langle A\rangle_{{\cal E}_+} - w_-\langle A\rangle_{{\cal E}_-},
\end{align}
where $w_\pm=\half(1\pm\delta)$ is determined (nonlinearly) in terms of conditional expectations via the solution of Eqs.~(\ref{l1})--(\ref{l3}). Similar rewriting of the remaining terms of the CHSH inequality then yields the equivalent nonlinear device independent steering inequality
\begin{align}
\big|w_+(&\langle A\rangle_{{\cal E}_+} +\langle B\rangle_{{\cal E}_+}) - w_-\left(\langle A\rangle_{{\cal E}_-} +\langle B\rangle_{{\cal E}_-}\right) \nn \\ 
&+ w'_+(\langle A\rangle_{{\cal E}'_+} -\langle B\rangle_{{\cal E}'_+}) - w'_-(\langle A\rangle_{{\cal E}'_-} -\langle B\rangle_{{\cal E}'_-})\big| \leq 2 ,
\end{align}
 with $w_\pm,w'_\pm$ determined via Eqs.~(\ref{l1})--(\ref{l3}). As noted in Sec.~\ref{sec:bellsteer}, it would be of interest to determine the conditions under which Bell inequalities for more general scenarios can be rewritten as device independent steering inequalities.
 
 Finally,  analogous results can be obtained via Lemmas~1 and~2 by replacing the locality assumption with operational completeness.

\section{Proof of Theorem~4}
\label{appth4}

Inequality~(\ref{iff}) of Theorem~4 corresponds to Pusey's necessary and sufficient condition in Sec.~VI of~\cite{pusey18}, for the compatibility of preparation noncontextuality with an ontological model, for measurements $A$ and $B$ and ensembles ${\cal E}_1, {\cal E}_2, {\cal E}_3, {\cal E}_4$ (note that our labelling  of ensembles  ${\cal E}_1, {\cal E}_2, {\cal E}_3, {\cal E}_4$ corresponds to the labelling ${\cal P}_0, {\cal P}_1, {\cal P}_3, {\cal P}_2$  of their respective preparation procedures  in~\cite{pusey18}, and we have reordered the bottom two rows of the determinant in Eq.~(11) of~\cite{pusey18}, thus changing the sign of the inequality). Pusey's condition in turn arises from a formal equivalence between preparation noncontextuality and local causal models for the CHSH scenario, as shown in Sec.~V of~\cite{pusey18}. Theorem~4 can therefore be established if a similar equivalence between operational completeness and local causal models can  be shown for this scenario, which we do as follows.

First, we show that if operational completeness and joint reality both hold, then  any four ensembles in an operational plane as per the statement of the theorem must satisfy Eq.~(\ref{iff}). It is notationally convenient to relabel the four subensembles, as
\beq \label{notation}
{\cal E}_1\equiv{\cal E}_+,~~~ {\cal E}_2\equiv{\cal E}_+',~~~ {\cal E}_3\equiv{\cal E}_-,~~~ {\cal E}_4\equiv{\cal E}_-' .
\eeq
Since they lie on an operational plane of $A$ and $B$, which by definition is closed under mixtures, and since the diagonals of a convex quadrilateral must intersect, it follows one can form a mixed ensemble ${\cal E}_/$  of ${\cal E}_+, {\cal E}_-$ with respective mixing fractions $w_+$, $w_-=1-w_+\in [0,1]$, and a mixed ensemble ${\cal E}_\backslash$  of  ${\cal E}_+', {\cal E}_-'$ with respective mixing fractions $w'_+$, $w'_-=1-w_+'\in [0,1]$, such that
\begin{align} \label{pm}
\langle A\rangle_{{\cal E}_/}\approx \langle A\rangle_{{\cal E}_\backslash}, \qquad\langle B\rangle_{{\cal E}_/}\approx \langle B\rangle_{{\cal E}_\backslash},
\end{align}
where $\langle C\rangle_{{\cal E}_/}= w_+ \langle C\rangle_{{\cal E}_+}+w_-\langle C\rangle_{{\cal E}_-}$ and 
$\langle C\rangle_{{\cal E}_\backslash}=w_+' \langle C\rangle_{{\cal E}_+'}+w_-'\langle C\rangle_{{\cal E}_-'}$ for $C=A,B$.
For example, for the ensembles represented by the red and blue dots in Fig.~\ref{semidi}, ${\cal E}_/$ and ${\cal E}_\backslash$ correspond to the black dot. 

Further, since $A$ and $B$ have real pre-existing values, then by construction their joint relative frequencies for ${\cal E}_/, {\cal E}_\backslash$ are given by
\beq
\frac{N(\alpha,\beta|{\cal E}_/)}{N_/}= w_+\frac{N(\alpha,\beta|{\cal E}_+)}{N_+}+ w_-  \frac{N(\alpha,\beta|{\cal E}_-)}{N_-} , 
\eeq
\beq
\frac{N(\alpha,\beta|{\cal E}_\backslash)}{N_\backslash}= w_+'\frac{N(\alpha,\beta|{\cal E}_+')}{N_+'}+ w_-'  \frac{N(\alpha,\beta|{\cal E}_-')}{N_-'},
\eeq
where $N_/, N_\backslash, N_\pm, N'_\pm$ denote the number of systems in ensembles ${\cal E}_/, {\cal E}_\backslash, {\cal E}_\pm, {\cal E}_\pm'$, respectively. It immediately follows via Eq.~(\ref{pm}) and operational completeness that 
\beq \label{curlyp}
\wp(\alpha,\beta):=\frac{N(\alpha,\beta|{\cal E}_/)}{N_/}\approx \frac{N(\alpha,\beta|{\cal E}_\backslash)}{N_\backslash} ,
\eeq  
up to statistical errors that become negligble for sufficiently large ensembles.

Consider now, guided by~\cite{pusey18}, a formal hidden variable model with hidden variables $u,v=\pm1$, for the joint statistics of measurements $C=A$ or $B$ and $D=M$ or $M'$, with respective measurement outcomes $\alpha,\beta,m,m'=\pm1$, of the form
\beq \label{lcm}
p(c,d|C,D) = \sum_{u,v} \wp(u,v)\, p_1(c|C,u,v)\,p_2(d|D,u,v).
\eeq
  We set  $p_1(\alpha|A,u,v):= \delta_{\alpha u}$, $p_1(\beta|B,u,v):= \delta_{\beta v}$, and
\beq
p_2(\pm|M,u,v):=\frac{w_\pm N(u,v|{\cal E}_\pm)/N_\pm}{N(u,v|{\cal E}_/)/N_/},
\eeq
\beq
p_2(\pm|M',u,v):=\frac{w_\pm' N(u,v|{\cal E}_\pm')/N_\pm'}{N(u,v|{\cal E}_\backslash)/N_\backslash} .
\eeq
Using Eq.~(\ref{curlyp}) and these definitions, Eq.~(\ref{lcm})  then  simplifies to
\begin{align}
p(c,\pm|C,M) &= w_\pm \frac{N(C=c|{\cal E}_\pm)}{N_\pm} = w_\pm \frac{1+c\langle C\rangle_{{\cal E}_\pm}}{2},\\
p(c,\pm|C,M') &\approx w_\pm' \frac{N(C=c|{\cal E}_\pm')}{N_\pm'} = w_\pm' \frac{1+c\langle C\rangle_{{\cal E}_\pm'}}{2},
\end{align}
where the final equalities follow from the definition of $\langle A\rangle$ and $\langle B\rangle$ for ensembles ${\cal E}_\pm, {\cal E}_\pm'$. Thus,  as far as observables $A$ and $B$ are concerned,  measurement of $M=\pm1$ in this formal model is equivalent to preparing ensemble ${\cal E}_\pm$ with probability $w_\pm$, while measurement of $M'=\pm1$ is equivalent, up to statistical errors, to preparing ensemble ${\cal E}_\pm'$ with probability $w_\pm'$. 

The formal hidden variable model above is constructed using the joint reality of $A$ and $B$, operational completeness, and ensembles ${\cal E}_1, {\cal E}_2, {\cal E}_3, {\cal E}_4$ as per the statement of the theorem [related to ${\cal E}_\pm, {\cal E}_\pm'$ via Eq.~(\ref{notation}) above]. Further, the form of Eq.~(\ref{lcm}) satisfies the requirements of local causality~\cite{bellreview}, for the case of two measurements on each side each having two outcomes. It follows that Eq.~(\ref{iff}) of the Theorem is necessarily satisfied by this formal model, since, as per the nontrivial result given in Sec.~VI of~\cite{pusey18}, this equation is equivalent to the eight CHSH inequalities satisfied by all such local causal models . This proves the theorem in the `if' direction.

Conversely, the `only if' direction follows if it can be shown that if Eq.~(\ref{iff}) is satisfied, for two observables $A$ and $B$ and four ensembles ${\cal E}_1, {\cal E}_2, {\cal E}_3, {\cal E}_4$ as per the statement of the theorem, then the joint reality of $A$ and $B$ is compatible with operational completeness. It turns out that we can show this even when the ensembles are not restricted to lie in some operational plane of $A$ and $B$, which is in fact a slightly stronger result. 

In particular, as shown in~\cite{pusey18}, Eq.~(\ref{iff}) is equivalent to the eight CHSH inequalities being satisfied by the joint statistics corresponding to  measurements $A$ and $B$ on one side and two two-valued measurements $M=\pm1$ and $M'=\pm1$ on the other, where the ensembles ${\cal E}_\pm, {\cal E}_\pm'$ correspond to the measurement outcomes of $M$ and $M'$, respectively, and ${\cal E}_+, {\cal E}_+', {\cal E}_-, {\cal E}_-'$ form a convex quadrilateral (with vertices labelled clockwise) in the $\langle A\rangle\langle B\rangle$-plane [note the correspondence in Eq.~(\ref{notation}) above].  Further, it is well known that when these CHSH inequalities are satisfied one can always construct a deterministic local causal model for the joint statistics, with a discrete hidden variable $\lambda$~\cite{fineprl}, i.e., of the form 
\beq \label{det}
\tilde p(c,d|C,D) = \sum_\lambda \wp(\lambda)\,  \tilde p_1(c|C,\lambda)\,\tilde p_2(d|D,\lambda),
\eeq
with $C=A,B$, $D=M,M'$, and $\tilde p_1(c|C,\lambda), \tilde p_2(d|D,\lambda)\in\{0,1\}$. Hence, each member of an ensemble $\cal E$ of $N$ systems described by these statistics is compatible with pre-existing values of $C=A,B$ and $D=M, M'$, specified by the deterministic joint distribution 
\begin{align}
&p(\alpha,\beta,m,m'|\lambda)\nn\\ \label{detjoint}
&~~~~~=\tilde p_1(\alpha|A,\lambda)\,  \tilde p_1(\beta|B,\lambda)\,\tilde p_2(m|M,\lambda)\,\tilde p_2(m'|M',\lambda).
\end{align} 
Thus, the model is compatible with the joint reality of $A$ and $B$ by construction. 

Now, local causality of the model in Eq.~(\ref{det}) implies that the joint statistics of $A$ and $B$ do not depend on whether $M$ or $M'$ is measured, i.e.,
\begin{align} 
\sum_{m} &p(\alpha,\beta|M=m)\,p(M=m)\nn\\
&= \sum_{\lambda, m,m'} \wp(\lambda)\, p(\alpha,\beta,m,m'|\lambda)\nn\\
&= \sum_{m'} p(\alpha,\beta|M'=m')\,p(M'=m'),
\end{align}
using Eqs.~(\ref{det}) and~(\ref{detjoint}). Hence, the subensembles corresponding to the measurement outcomes for $M$ and $M'$ satisfy
\begin{align}
&\sum_m p(M=m)\frac{N(\alpha,\beta|{\cal E}_m)}{N_m}\approx 
\sum_{m'} p(M'=m')\frac{N(\alpha,\beta|{\cal E}_{m'}')}{N_{m'}'},
\end{align}
where $N_\pm, N_\pm'$ denote the number of systems in subensembles ${\cal E}_\pm, {\cal E}_\pm'$ as before, up to statistical errors that become negligible as size of the subensemble is increased. Identifying the left hand side of this equation with a mixed ensemble ${\cal E}_/$ of  ${\cal E}_\pm$, and the right hand side with a mixed ensemble ${\cal E}_\backslash$ of ${\cal E}_\pm'$, then gives
\beq
\frac{N(\alpha,\beta|{\cal E}_/)}{N_/}\approx \frac{N(\alpha,\beta|{\cal E}_\backslash)}{N_\backslash}
\eeq
[analogous to Eq.~(\ref{curlyp})]. Thus, the joint relative frequencies of $A$ and $B$ (and hence the averages of $A$ and $B$), are equal up to statistical errors, implying that ${\cal E}_/$ and ${\cal E}_\backslash$ satisfy the condition required for operational completeness in Sec.~\ref{sec:gen} (whether or not they are operationally similar, i.e, even if the subensembles ${\cal E}_\pm, {\cal E}_\pm'$ do not lie in an operational plane). Finally, no other points on the $\langle A\rangle\langle B\rangle$-plane need to be considered for the purposes of operational completeness, since the diagonals of a convex quadrilateral in this plane only cross at a single point.  Hence the joint reality of $A$ and $B$ is compatible with operational completeness,  as required for the `only if' direction of Theorem~4.


\begin{thebibliography}{99}
\bibitem{footex} This example is a minor variant of the model given by Bell in Sec.~V of~\cite{bellrmp}. It may be checked that averaging over $\lambda$ yields the correct quantum prediction $p(A=\alpha|\psi)=\half(1+ \alpha  \langle A\rangle_\psi)$ for $A$, and likewise for $B$.
\bibitem{probid} This identity is most easily confirmed by checking that it solves the four linear equations  $\langle A\rangle=N^{-1} \sum_{\alpha,\beta} \alpha N(\alpha,\beta)$,  $\langle B\rangle=N^{-1} \sum_{\alpha,\beta} \beta N(\alpha,\beta)$,  $\langle AB\rangle=N^{-1} \sum_{\alpha,\beta} \alpha\beta N(\alpha,\beta)$, $1=N^{-1}  \sum_{\alpha,\beta} N(\alpha,\beta)$.
\bibitem{mnc} Similarly, an ontological model is  defined to be measurement noncontextual if any two operationally equivalent measurements are  equivalent at the ontic level~\cite{spek05}. However, measurement noncontextuality, whether in this form or any other~\cite{ks,other}, is not relevant here and so will not be discussed further (in particular, to avoid any implicit assumption of measurement noncontextuality in Sec.~\ref{sec:corr} we can suppose that a single fixed method of measurement has been chosen for each observable).
\end{thebibliography}
\end{document}